\documentstyle[12pt,a4]{article}

\begin{document}
\setlength{\oddsidemargin}{-1cm}
\setlength{\evensidemargin}{-1cm}
\newcommand{\nc}{\newcommand}
\nc{\beq}{\begin{equation}}
\nc{\eeq}{\end{equation}}
\nc{\bea}{\begin{eqnarray}}
\nc{\eea}{\end{eqnarray}}
\nc{\bpi}{\begin{picture}}
\nc{\epi}{\end{picture}}
\nc{\ba}{\begin{array}}
\nc{\ea}{\end{array}}
\nc{\nn}{\nonumber}
\nc{\ts}{\textstyle}
\nc{\ds}{\displaystyle}
\nc{\sss}{\scriptscriptstyle}
\nc{\bm}{\boldmath}
\nc{\Int}[1]{{\ds\int_{#1}}}
\nc{\mubar}{\bar{\mu}}
\nc{\fm}{\frac{m^2}{\mubar^2}}
\nc{\lnma}{\ln\fm}
\nc{\lnmb}{\ln^2\fm}
\nc{\lnmc}{\ln^3\fm}
\nc{\lnmd}{\ln^4\fm}
\nc{\msbar}{$\overline{\mbox{\rm MS}}$}
\nc{\od}{{\cal O}}
\nc{\al}{\alpha}
\nc{\be}{\beta}
\nc{\ga}{\gamma}
\nc{\de}{\delta}
\nc{\ep}{\epsilon}
\nc{\ze}{\zeta}
\nc{\p}{\partial}

\begin{flushright}
\normalsize
Freiburg-THEP 97/22\\
October 1997\vspace{0.8cm}\\
\end{flushright}

\begin{center}
{\large\bf\bm  Five-Loop Vacuum Energy $\be$ Function in $\phi^4$ Theory
with $O(N)$-Symmetric and Cubic Interactions\vspace{1.2cm}\\}
Boris Kastening\vspace{0.4cm}\\
\it Albert-Ludwigs-Universit\"at Freiburg\\
\it Fakult\"at f\"ur Physik\\
\it Hermann-Herder-Stra\ss e 3\\
\it D-79104 Freiburg\\
\it Germany\vspace{1.2cm}\\
\end{center}

\begin{center}
{\bf Abstract}\vspace{0.4cm}\\
\begin{minipage}{15cm}
The beta function of the vacuum energy density is analytically computed
at the five-loop level in $O(N)$-symmetric $\phi^4$ theory, using
dimensional regularization in conjunction with the \msbar\ scheme.
The result for the case of cubic anisotropy is also given.
It is pointed out how to also obtain the beta function of the
coupling and the gamma function of the mass from vacuum graphs.
This method may be easier than traditional approaches.
\end{minipage}
\vspace{1.2cm}\\
\end{center}

\section{Introduction}
In this article, we extend earlier work \cite{Ka}, where the beta
function $\be_v$ of the vacuum energy density in $O(N)$-symmetric
$\phi^4$ theory was computed at the four-loop level, to five loops.
Integrals are dimensionally regulated and divergences removed by
modified minimal subtraction (\msbar).
For motivations for this work, see \cite{Ka}.
We employ two different methods to arrive at our result for the
$O(N)$-symmetric case.
In section \ref{myscheme}, we use the scheme from \cite{Ka} to determine
the five-loop contributions to the vacuum energy renormalization constant
$Z_v$ and to $\be_v$.
All necessary subtractions result from consistency requirements while
renormalizing the vacuum energy.
As explained in \cite{Ka}, this yields as a by-product $\ga_m$
through four loops and $\be_g$ through three loops.
In section \ref{recursion}, we check the relevant recursion relations
for the contributions to $Z_v$.
In section \ref{betav}, we give $\be_v$ through five loops for the
$O(N)$-symmetric case.
In section \ref{standard}, we use a more traditional approach to check
our result for $\be_v$.
In section \ref{cubic}, we extend it for the case of an additional
cubic interaction.
Our work brings the evaluation of $\be_v$ on a par with that
of the other beta and gamma functions in $\phi^4$ theory
(see \cite{Kl..,KlSc} and references therein).

For definitions and conventions, the reader is referred to the detailed
article in \cite{Ka}.
The only exception is $\be_v$ itself, which we define here as
\beq
\label{betavdef}
\be_{v,\ep}(g,\ep)
=\frac{\mu^{2+\ep}}{m^4}
\left[\frac{\p}{\p\mu^{2}}
\left(\frac{m^4 h}{\mu^\ep g}\right)\right]_{B}\,,
\eeq
where the subscript $B$ indicates that bare quantities are kept fixed.
I.e.\ $\be_v$ and $\be_{v,\ep}$ in this article are just $\be_v$
and $\be_{v,\ep}$ from \cite{Ka} divided by the renormalized coupling $g$.
The connection to the constant
$Z_v=1+h^{-1}\sum_{k=1}^\infty Z_{v,k}(g)/\ep^k$, which renormalizes
the vacuum energy, is now
\beq
\label{betavfromz}
\be_{v,\ep}(g,\ep)=\be_v(g)=\frac{1}{2}Z_{v,1}'\,.
\eeq
Our revised definition is more natural since even in the non-interacting
theory, i.e.\ for $g=0$, the vacuum energy density acquires a divergent part
from the zero-point fluctuations of the uncoupled harmonic oscillators.
It has to be renormalized and therefore runs with $\mu$, such that
\beq
\label{betav0}
\be_v=\frac{N}{4}
\eeq
for the free theory.
E.g., the renormalization group equation for the effective potential
in $d=4-\ep$ dimensions for $h=0$ reads now \cite{Ka}
\beq
\label{rge}
\left[
\mu^2\frac{\p}{\p\mu^2}
+\be_{g,\ep}\frac{\p}{\p g}
+\ga_{m,\ep}m^2\frac{\p}{\p m^2}
-\ga_{\phi,\ep}\phi^2\frac{\p}{\p \phi^2}
\right]V_\ep(g,m^2,\phi^2,h=0,\mu^2)
=
-\frac{\be_{v,\ep}m^4}{(4\pi)^2\mu^\ep}\,.
\eeq
For the free theory, we have $\be_{g,\ep}=\ga_{m,\ep}=\ga_{\phi,\ep}=0$
and, going to four dimensions, the \msbar\ vacuum energy
\beq
V(g=0,m^2,\phi^2=0,h=0,\mu^2)=\frac{Nm^4}{4(4\pi)^2}
\left(\lnma-\frac{3}{2}\right)\,,
\eeq
with the \msbar\ renormalization scale
\beq
\mubar^2\equiv 4\pi\mu^2 e^{-\ga_E}
\eeq
is immediately seen to satisfy (\ref{rge}) with $\be_v$ given by
(\ref{betav0}) with no spurious appearances of a coupling.

\section{\bm Five-Loop Contribution to $Z_v$ and $\be_v$}
\label{myscheme}
In this section, we determine the five-loop contributions to $Z_v$
and to $\be_v$ in the $O(N)$-symmetric model.
We proceed in the same way as in \cite{Ka}:
We employ modified Feynman rules, where the one-loop mass correction
is absorbed into a modified mass in the propagator.
\begin{table}[t]
\begin{center}
\begin{tabular}{c|c|c}
\hspace{-10pt}
\begin{tabular}{c}number\\of loops\end{tabular}
\hspace{-10pt}
&
\hspace{-10pt}
\begin{tabular}{c}order\\in $g$\end{tabular}
\hspace{-10pt}
&
diagrams and symmetry factors
\\
\hline
0&$g^{-1}\!\!\!\!$&
$1$
\hspace{-10pt}
\rule[-2pt]{0pt}{10pt}
\bpi(10,12)
\put(5,3){\circle*{4}}
\epi
\\
\hline
1&$g^0$&
$\frac{1}{2}$
\hspace{-10pt}
\rule[-10pt]{0pt}{26pt}
\bpi(26,12)
\put(13,3){\circle{16}}
\epi
\\
\hline
2&$g^1$&
$\frac{1}{8}$
\hspace{-10pt}
\rule[-10pt]{0pt}{26pt}
\bpi(42,12)
\put(13,3){\circle{16}}
\put(29,3){\circle{16}}
\put(21,3){\circle*{4}}
\epi
\\
\hline
3&$g^2$&
$\!\frac{1}{48}$
\hspace{-10pt}
\rule[-14pt]{0pt}{34pt}
\bpi(34,12)
\put(17,3){\circle{24}}
\put(17,3){\oval(24,8)}
\put(5,3){\circle*{4}}
\put(29,3){\circle*{4}}
\epi
\hspace{10pt}
$\frac{1}{16}$
\hspace{-10pt}
\rule[-10pt]{0pt}{26pt}
\bpi(58,12)
\put(13,3){\circle{16}}
\put(29,3){\circle{16}}
\put(45,3){\circle{16}}
\put(21,3){\circle*{4}}
\put(37,3){\circle*{4}}
\epi
\\
\hline
4&$g^3$&
$\!\frac{1}{48}$
\hspace{-10pt}
\rule[-14pt]{0pt}{34pt}
\bpi(34,12)
\put(17,3){\circle{24}}
\put(6.6,9){\line(1,0){20.8}}
\put(6.6,9){\line(3,-5){10.4}}
\put(27.4,9){\line(-3,-5){10.4}}
\put(6.6,9){\circle*{4}}
\put(27.4,9){\circle*{4}}
\put(17,-9){\circle*{4}}
\epi
\hspace{10pt}
$\!\frac{1}{24}$
\hspace{-10pt}
\rule[-14pt]{0pt}{50pt}
\bpi(34,12)
\put(17,3){\circle{24}}
\put(17,3){\oval(24,8)}
\put(17,23){\circle{16}}
\put(5,3){\circle*{4}}
\put(29,3){\circle*{4}}
\put(17,15){\circle*{4}}
\epi
\hspace{10pt}
$\frac{1}{32}$
\hspace{-10pt}
\rule[-10pt]{0pt}{26pt}
\bpi(74,12)
\put(13,3){\circle{16}}
\put(29,3){\circle{16}}
\put(45,3){\circle{16}}
\put(61,3){\circle{16}}
\put(21,3){\circle*{4}}
\put(37,3){\circle*{4}}
\put(53,3){\circle*{4}}
\epi
\hspace{10pt}
$\!\frac{1}{48}$
\hspace{-10pt}
\rule[-18pt]{0pt}{50pt}
\bpi(53.7,12)
\put(26.85,3){\circle{16}}
\put(26.85,19){\circle{16}}
\put(13,-5){\circle{16}}
\put(40.7,-5){\circle{16}}
\put(26.85,11){\circle*{4}}
\put(19.95,-1){\circle*{4}}
\put(33.75,-1){\circle*{4}}
\epi
\\
\hline
5&$g^4$&
$\frac{1}{128}$
\hspace{-10pt}
\rule[-14pt]{0pt}{34pt}
\bpi(34,12)
\put(17,3){\circle{24}}
\put(8.5,-5.5){\line(1,0){17}}
\put(8.5,-5.5){\line(0,1){17}}
\put(8.5,11.5){\line(1,0){17}}
\put(25.5,-5.5){\line(0,1){17}}
\put(8.5,-5.5){\circle*{4}}
\put(8.5,11.5){\circle*{4}}
\put(25.5,-5.5){\circle*{4}}
\put(25.5,11.5){\circle*{4}}
\epi
\hspace{10pt}
$\frac{1}{144}$
\hspace{-10pt}
\rule[-22pt]{0pt}{50pt}
\bpi(42,12)
\put(21,-9){\circle{16}}
\put(21,15){\circle{16}}
\put(13,-9){\line(1,0){16}}
\put(13,15){\line(1,0){16}}
\put(13,3){\oval(16,24)[l]}
\put(29,3){\oval(16,24)[r]}
\put(13,-9){\circle*{4}}
\put(13,15){\circle*{4}}
\put(29,-9){\circle*{4}}
\put(29,15){\circle*{4}}
\epi
\hspace{10pt}
$\frac{1}{32}$
\hspace{-10pt}
\rule[-18pt]{0pt}{42pt}
\bpi(58,12)
\put(13,3){\circle{16}}
\put(45,3){\circle{16}}
\put(5,-5){\line(0,1){16}}
\put(25,-5){\oval(40,16)[b]}
\put(25,11){\oval(40,16)[t]}
\put(45,3){\oval(48,16)[l]}
\put(5,3){\circle*{4}}
\put(21,3){\circle*{4}}
\put(45,-5){\circle*{4}}
\put(45,11){\circle*{4}}
\epi
\hspace{10pt}
$\frac{1}{16}$
\hspace{-10pt}
\rule[-14pt]{0pt}{50pt}
\bpi(34,12)
\put(17,3){\circle{24}}
\put(17,23){\circle{16}}
\put(6.6,9){\line(1,0){20.8}}
\put(6.6,9){\line(3,-5){10.4}}
\put(27.4,9){\line(-3,-5){10.4}}
\put(6.6,9){\circle*{4}}
\put(27.4,9){\circle*{4}}
\put(17,-9){\circle*{4}}
\put(17,15){\circle*{4}}
\epi
\hspace{10pt}
$\frac{1}{48}$
\hspace{-10pt}
\rule[-14pt]{0pt}{47.3pt}
\bpi(46,12)
\put(23,3){\circle{24}}
\put(13,20.3){\circle{16}}
\put(33,20.3){\circle{16}}
\put(23,3){\oval(24,8)}
\put(11,3){\circle*{4}}
\put(35,3){\circle*{4}}
\put(17,13.4){\circle*{4}}
\put(29,13.4){\circle*{4}}
\epi
\hspace{10pt}
$\frac{1}{32}$
\hspace{-10pt}
\rule[-30pt]{0pt}{66pt}
\bpi(34,12)
\put(17,3){\circle{24}}
\put(17,-17){\circle{16}}
\put(17,23){\circle{16}}
\put(17,3){\oval(24,8)}
\put(5,3){\circle*{4}}
\put(17,-9){\circle*{4}}
\put(17,15){\circle*{4}}
\put(29,3){\circle*{4}}
\epi
\\
&&
$\frac{1}{48}$
\hspace{-10pt}
\rule[-14pt]{0pt}{66pt}
\bpi(34,12)
\put(17,3){\circle{24}}
\put(17,23){\circle{16}}
\put(17,39){\circle{16}}
\put(17,3){\oval(24,8)}
\put(5,3){\circle*{4}}
\put(17,15){\circle*{4}}
\put(17,31){\circle*{4}}
\put(29,3){\circle*{4}}
\epi
\hspace{10pt}
$\frac{1}{64}$
\hspace{-10pt}
\rule[-10pt]{0pt}{26pt}
\bpi(90,12)
\put(13,3){\circle{16}}
\put(29,3){\circle{16}}
\put(45,3){\circle{16}}
\put(61,3){\circle{16}}
\put(77,3){\circle{16}}
\put(21,3){\circle*{4}}
\put(37,3){\circle*{4}}
\put(53,3){\circle*{4}}
\put(69,3){\circle*{4}}
\epi
\hspace{10pt}
$\frac{1}{32}$
\hspace{-10pt}
\rule[-23.85pt]{0pt}{53.7pt}
\bpi(66,12)
\put(13,3){\circle{16}}
\put(29,3){\circle{16}}
\put(45,3){\circle{16}}
\put(53,-10.85){\circle{16}}
\put(53,16.85){\circle{16}}
\put(21,3){\circle*{4}}
\put(37,3){\circle*{4}}
\put(49,-3.9){\circle*{4}}
\put(49,9.9){\circle*{4}}
\epi
\hspace{10pt}
$\frac{1}{128}$
\hspace{-10pt}
\rule[-26pt]{0pt}{58pt}
\bpi(58,12)
\put(13,3){\circle{16}}
\put(29,-13){\circle{16}}
\put(29,3){\circle{16}}
\put(29,19){\circle{16}}
\put(45,3){\circle{16}}
\put(21,3){\circle*{4}}
\put(29,-5){\circle*{4}}
\put(29,11){\circle*{4}}
\put(37,3){\circle*{4}}
\epi
\end{tabular}
\end{center}
\caption{\protect\label{allgraphs}
Vacuum diagrams through five loops and their symmetry factors.
In the equations in the text the symmetry factor is considered part of
each respective diagram.}
\end{table}
\begin{table}[t]
\begin{center}
\begin{tabular}{c|c|ccc}
\begin{tabular}{c}number\\of loops\end{tabular}
&
\begin{tabular}{c}order\\in $g$\end{tabular}
&
\multicolumn{3}{c}{
\begin{tabular}{c}remaining diagrams and\\
revised symmetry factors\end{tabular}}
\\
\hline
0&$g^{-1}\!\!\!\!$&
$1$
\rule[-2pt]{0pt}{10pt}
\bpi(10,12)
\put(5,3){\circle*{4}}
\epi
\\
\hline
1&$g^0$&
$\frac{1}{2}$
\rule[-10pt]{0pt}{26pt}
\bpi(26,12)
\put(13,3){\circle{16}}
\epi
\\
\hline
2&$g^1$&
$-\frac{1}{8}$
\rule[-10pt]{0pt}{26pt}
\bpi(42,12)
\put(13,3){\circle{16}}
\put(29,3){\circle{16}}
\put(21,3){\circle*{4}}
\epi
\\
\hline
3&$g^2$&
$\frac{1}{48}$
\rule[-14pt]{0pt}{34pt}
\bpi(34,12)
\put(17,3){\circle{24}}
\put(17,3){\oval(24,8)}
\put(5,3){\circle*{4}}
\put(29,3){\circle*{4}}
\epi
\\
\hline
4&$g^3$&
$\frac{1}{48}$
\rule[-14pt]{0pt}{34pt}
\bpi(34,12)
\put(17,3){\circle{24}}
\put(6.6,9){\line(1,0){20.8}}
\put(6.6,9){\line(3,-5){10.4}}
\put(27.4,9){\line(-3,-5){10.4}}
\put(6.6,9){\circle*{4}}
\put(27.4,9){\circle*{4}}
\put(17,-9){\circle*{4}}
\epi
\\
\hline
5&$g^4$&
$\frac{1}{128}$
\rule[-14pt]{0pt}{34pt}
\bpi(34,12)
\put(17,3){\circle{24}}
\put(8.5,-5.5){\line(1,0){17}}
\put(8.5,-5.5){\line(0,1){17}}
\put(8.5,11.5){\line(1,0){17}}
\put(25.5,-5.5){\line(0,1){17}}
\put(8.5,-5.5){\circle*{4}}
\put(8.5,11.5){\circle*{4}}
\put(25.5,-5.5){\circle*{4}}
\put(25.5,11.5){\circle*{4}}
\epi
&
$\frac{1}{144}$
\rule[-22pt]{0pt}{50pt}
\bpi(42,12)
\put(21,-9){\circle{16}}
\put(21,15){\circle{16}}
\put(13,-9){\line(1,0){16}}
\put(13,15){\line(1,0){16}}
\put(13,3){\oval(16,24)[l]}
\put(29,3){\oval(16,24)[r]}
\put(13,-9){\circle*{4}}
\put(13,15){\circle*{4}}
\put(29,-9){\circle*{4}}
\put(29,15){\circle*{4}}
\epi
&
$\frac{1}{32}$
\rule[-18pt]{0pt}{42pt}
\bpi(58,12)
\put(13,3){\circle{16}}
\put(45,3){\circle{16}}
\put(5,-5){\line(0,1){16}}
\put(25,-5){\oval(40,16)[b]}
\put(25,11){\oval(40,16)[t]}
\put(45,3){\oval(48,16)[l]}
\put(5,3){\circle*{4}}
\put(21,3){\circle*{4}}
\put(45,-5){\circle*{4}}
\put(45,11){\circle*{4}}
\epi
\\
\hline
6&$g^5$&
$\frac{1}{320}$
\rule[-14pt]{0pt}{34pt}
\bpi(34,12)
\put(17,3){\circle{24}}
\put(17,15){\circle*{4}}
\put(5.59,6.71){\circle*{4}}
\put(28.41,6.71){\circle*{4}}
\put(9.95,-6.71){\circle*{4}}
\put(24.05,-6.71){\circle*{4}}
\put(9.95,-6.71){\line(1,0){14.1}}
\put(9.95,-6.71){\line(-1,3){4.36}}
\put(24.04,-6.71){\line(1,3){4.36}}
\put(5.59,6.71){\line(4,3){11.41}}
\put(28.41,6.71){\line(-4,3){11.41}}
\epi
&
$\frac{1}{288}$
\rule[-14pt]{0pt}{34pt}
\bpi(58,12)
\put(17,3){\circle{24}}
\put(41,3){\circle{24}}
\put(17,3){\oval(8,24)}
\put(41,3){\oval(8,24)}
\put(29,3){\circle*{4}}
\put(17,15){\circle*{4}}
\put(17,-9){\circle*{4}}
\put(41,15){\circle*{4}}
\put(41,-9){\circle*{4}}
\epi
&
$\frac{1}{48}$
\rule[-18pt]{0pt}{42pt}
\bpi(66,12)
\put(13,3){\circle{16}}
\put(13,-5){\line(0,1){16}}
\put(37,11){\oval(48,16)[t]}
\put(37,-5){\oval(48,16)[b]}
\put(45,-5){\circle{16}}
\put(45,11){\circle{16}}
\put(61,-5){\line(0,1){16}}
\put(13,11){\circle*{4}}
\put(13,-5){\circle*{4}}
\put(45,-13){\circle*{4}}
\put(45,3){\circle*{4}}
\put(45,19){\circle*{4}}
\epi
\\
\rule[-18pt]{0pt}{42pt}
&&
$\frac{1}{32}$
\bpi(74,12)
\put(13,3){\circle{16}}
\put(29,3){\circle{16}}
\put(61,3){\circle{16}}
\put(5,-5){\line(0,1){16}}
\put(33,-5){\oval(56,16)[b]}
\put(33,11){\oval(56,16)[t]}
\put(61,3){\oval(48,16)[l]}
\put(5,3){\circle*{4}}
\put(21,3){\circle*{4}}
\put(37,3){\circle*{4}}
\put(61,-5){\circle*{4}}
\put(61,11){\circle*{4}}
\epi
&
$\frac{1}{16}$
\rule[-18pt]{0pt}{42pt}
\bpi(74,12)
\put(13,3){\circle{16}}
\put(61,3){\circle{16}}
\put(13,3){\oval(48,16)[r]}
\put(61,3){\oval(48,16)[l]}
\put(37,11){\oval(48,16)[t]}
\put(37,-5){\oval(48,16)[b]}
\put(13,11){\circle*{4}}
\put(13,-5){\circle*{4}}
\put(37,3){\circle*{4}}
\put(61,11){\circle*{4}}
\put(61,-5){\circle*{4}}
\epi
&
$\frac{1}{120}$
\rule[-14pt]{0pt}{34pt}
\bpi(34,12)
\put(17,3){\circle{24}}
\put(17,15){\circle*{4}}
\put(5.59,6.71){\circle*{4}}
\put(28.41,6.71){\circle*{4}}
\put(9.95,-6.71){\circle*{4}}
\put(24.05,-6.71){\circle*{4}}
\put(5.59,6.71){\line(1,0){22.8}}
\put(9.95,-6.71){\line(1,3){7.05}}
\put(24.05,-6.71){\line(-1,3){7.05}}
\put(9.95,-6.71){\line(4,3){18.46}}
\put(24.05,-6.71){\line(-4,3){18.46}}
\epi
\end{tabular}
\end{center}
\caption{\protect\label{remaininggraphs}
Remaining diagrams through six-loop order after absorption of
the one-loop mass correction into a modified mass in the propagator, i.e.\
after a careful resummation of the quadratic part of the Lagrange density.}
\end{table}
This reduces the full set of diagrams in table \ref{allgraphs}
to the reduced set in table \ref{remaininggraphs}, where we have
also included the six-loop diagrams, whose evaluation we save for
a later day.
We arrange that the appearances of the wave function renormalization
constant $Z_\phi$ in the propagator and coupling cancel from the outset.
There are no extra counterterm rules, since all counterterms are already
contained in the Feynman rules for vacuum energy, propagator and
coupling; i.e.\ we use bare values for vacuum energy, mass and coupling
in the Feynman rules and expand our results for diagrams to the necessary
order in the renormalized coupling $g$.
For this purpose, we have to formally construct the renormalization
constants $Z_m$, $Z_{\bar{m}}$ and $Z_g$, for mass, modified mass and
coupling, through four, four and three loops, respectively, from
$\ga_m$ and $\be_g$ \cite{Ka}.

The calculation through four loops is detailed in \cite{Ka}.
Continuing to five loops, we keep all divergent terms in zero through
five loops through order $g^4$.
The relevant five-loop diagrams are
\bea
\rule[-14pt]{0pt}{34pt}
\bpi(34,12)
\put(17,3){\circle{24}}
\put(8.5,-5.5){\line(1,0){17}}
\put(8.5,-5.5){\line(0,1){17}}
\put(8.5,11.5){\line(1,0){17}}
\put(25.5,-5.5){\line(0,1){17}}
\put(8.5,-5.5){\circle*{4}}
\put(8.5,11.5){\circle*{4}}
\put(25.5,-5.5){\circle*{4}}
\put(25.5,11.5){\circle*{4}}
\epi
&=&
\frac{N(N+2)(N^2+6N+20)}{10368}(4\pi)^8g^4Z_g^4 Z_{\bar{m}}^{2-\frac{5}{2}\ep}
I_{5a}\,,
\\
\rule[-22pt]{0pt}{50pt}
\bpi(42,12)
\put(21,-9){\circle{16}}
\put(21,15){\circle{16}}
\put(13,-9){\line(1,0){16}}
\put(13,15){\line(1,0){16}}
\put(13,3){\oval(16,24)[l]}
\put(29,3){\oval(16,24)[r]}
\put(13,-9){\circle*{4}}
\put(13,15){\circle*{4}}
\put(29,-9){\circle*{4}}
\put(29,15){\circle*{4}}
\epi
&=&
\frac{N(N+2)^2}{1296}(4\pi)^8g^4Z_g^4 Z_{\bar{m}}^{2-\frac{5}{2}\ep}
I_{5b}\,,
\\
\rule[-18pt]{0pt}{42pt}
\bpi(58,12)
\put(13,3){\circle{16}}
\put(45,3){\circle{16}}
\put(5,-5){\line(0,1){16}}
\put(25,-5){\oval(40,16)[b]}
\put(25,11){\oval(40,16)[t]}
\put(45,3){\oval(48,16)[l]}
\put(5,3){\circle*{4}}
\put(21,3){\circle*{4}}
\put(45,-5){\circle*{4}}
\put(45,11){\circle*{4}}
\epi
&=&
\frac{N(N+2)(5N+22)}{2592}(4\pi)^8g^4Z_g^4 Z_{\bar{m}}^{2-\frac{5}{2}\ep}
I_{5c}\,,
\eea
where $I_{5a}$, $I_{5b}$ and $I_{5c}$ are defined, their evaluation
sketched and their results given in appendix \ref{five-loop integrals}.
Using these, the formally reconstructed $Z_m$, $Z_{\bar{m}}$ and $Z_g$,
as well as the results of \cite{Ka}, one gets
\bea
\label{vacuum energy}
\lefteqn{
\rule[-14pt]{0pt}{34pt}
\bpi(14,12)
\put(7,3){\circle*{4}}
\epi
+
\bpi(26,12)
\put(13,3){\circle{16}}
\epi
+
\bpi(42,12)
\put(13,3){\circle{16}}
\put(29,3){\circle{16}}
\put(21,3){\circle*{4}}
\epi
+
\bpi(34,12)
\put(17,3){\circle{24}}
\put(17,3){\oval(24,8)}
\put(5,3){\circle*{4}}
\put(29,3){\circle*{4}}
\epi
+
\bpi(34,12)
\put(17,3){\circle{24}}
\put(6.6,9){\line(1,0){20.8}}
\put(6.6,9){\line(3,-5){10.4}}
\put(27.4,9){\line(-3,-5){10.4}}
\put(6.6,9){\circle*{4}}
\put(27.4,9){\circle*{4}}
\put(17,-9){\circle*{4}}
\epi
+
\rule[-14pt]{0pt}{34pt}
\bpi(34,12)
\put(17,3){\circle{24}}
\put(8.5,-5.5){\line(1,0){17}}
\put(8.5,-5.5){\line(0,1){17}}
\put(8.5,11.5){\line(1,0){17}}
\put(25.5,-5.5){\line(0,1){17}}
\put(8.5,-5.5){\circle*{4}}
\put(8.5,11.5){\circle*{4}}
\put(25.5,-5.5){\circle*{4}}
\put(25.5,11.5){\circle*{4}}
\epi
+
\rule[-22pt]{0pt}{50pt}
\bpi(42,12)
\put(21,-9){\circle{16}}
\put(21,15){\circle{16}}
\put(13,-9){\line(1,0){16}}
\put(13,15){\line(1,0){16}}
\put(13,3){\oval(16,24)[l]}
\put(29,3){\oval(16,24)[r]}
\put(13,-9){\circle*{4}}
\put(13,15){\circle*{4}}
\put(29,-9){\circle*{4}}
\put(29,15){\circle*{4}}
\epi
+
\rule[-18pt]{0pt}{42pt}
\bpi(58,12)
\put(13,3){\circle{16}}
\put(45,3){\circle{16}}
\put(5,-5){\line(0,1){16}}
\put(25,-5){\oval(40,16)[b]}
\put(25,11){\oval(40,16)[t]}
\put(45,3){\oval(48,16)[l]}
\put(5,3){\circle*{4}}
\put(21,3){\circle*{4}}
\put(45,-5){\circle*{4}}
\put(45,11){\circle*{4}}
\epi
}
\nn\\
&=&\ts
-\frac{m^4}{(4\pi)^2g}\bigg\{h
\nn\\
&&\ts\hspace{20pt}
+\Big[Z_{15}^v+\frac{5N(N+2)}{144}\left(\be_3
-\frac{33N^2+922N+2960+96(5N+22)\ze(3))}{432}
\right)\left(\lnma-1\right)^2
\nn\\
&&\ts\hspace{30pt}
+\frac{N}{4}\left(\al_4-\frac{(N+2)[
N^2-7578N-31060-48(3N^2+10N+68)\ze(3)-288(5N+22)\ze(4)]}{15552}
\right)\left(\lnma-1\right)
\nn\\
&&\ts\hspace{55pt}
+\frac{N(N+2)}{622080}\Big(
627N^2-51124N-224192+4320\be_3
\nn\\
&&\ts\hspace{105pt}
+4320\left(\be_3
-\frac{33N^2+922N+2960+96(5N+22)\ze(3)}{432}\right)\ze(2)
\nn\\
&&\ts\hspace{105pt}
-(144N^2-13536N-60480)\ze(3)
-(1152N^2+11232N+42048)\zeta(4)
\nn\\
&&\ts\hspace{105pt}
+(15360N+67584)\zeta(5)
\Big)\Big]
\frac{g^5}{\ep}
\nn\\
&&\ts\hspace{20pt}
+\Big[Z_{25}^v-\frac{N(N+2)}{36}\left(\be_3
-\frac{33N^2+922N+2960+96(5N+22)\ze(3)}{432}
\right)\left(\lnma-1\right)
\nn\\
&&\ts\hspace{30pt}
-\frac{N}{2}\left(\al_4+\frac{(N+2)[
-501N^2+226N+6404+96(4N^2+39N+146)\ze(3)+288(5N+22)\ze(4)]}
{77760}\right)\Big]
\frac{g^5}{\ep^2}
\nn\\
&&\ts\hspace{20pt}
+\left[Z_{35}^v-\frac{N(N+2)}{18}\left(\be_3+
\frac{873N^2+12314N+35296+96(5N+22)\ze(3)}{2160}\right)\right]
\frac{g^5}{\ep^3}
\nn\\
&&\ts\hspace{20pt}
+\left[Z_{45}^v+\frac{N(N+2)(293N^2+2624N+5840)}{4860}\right]
\frac{g^5}{\ep^4}
\nn\\
&&\ts\hspace{20pt}
+\left[Z_{55}^v-\frac{N(N+2)(N+4)(N+5)(5N+28)}{810}\right]
\frac{g^5}{\ep^5}
+\od(g^6,\ep^0)\bigg\}\,.
\eea
Demanding the cancellation of logarithmic terms gives the three-loop
coefficient of $\be_g$ and the four-loop coefficient of $\ga_m$,
\bea
\label{beta3}
\be_3
&=&\ts
\frac{33N^2+922N+2960+96(5N+22)\ze(3)}{432}\,,
\\
\label{alpha4}
\ts\al_4
&=&\ts
\frac{(N+2)[
N^2-7578N-31060-48(3N^2+10N+68)\ze(3)-288(5N+22)\ze(4)]}{15552}\,,
\eea
which coincide with known results (see, e.g., \cite{Kl..,KlSc}).
Demanding subsequently (\ref{vacuum energy}) to be finite as
$\ep\rightarrow 0$ gives
\bea
\label{z15z25z35z45z55}
Z_{15}^v
&=&\ts
\frac{N(N+2)
[-319N^2+13968N+64864+16(3N^2-382N-1700)\ze(3)+96(4N^2+39N+146)\ze(4)
-1024(5N+22)\ze(5)]}{207360}\,,
\\
Z_{25}^v
&=&\ts
-\frac{N(N+2)
[31N^2+2354N+9306+3(7N^2-28N+48)\ze(3)+72(5N+22)\ze(4)]}{9720}\,,
\\
Z_{35}^v
&=&\ts
\frac{N(N+2)[519N^2+8462N+25048+288(5N+22)\ze(3)]}{19440}\,,
\\
Z_{45}^v
&=&\ts
-\frac{N(N+2)(293N^2+2624N+5840)}{4860}\,,
\\
Z_{55}^v
&=&\ts
\frac{N(N+2)(N+4)(N+5)(5N+28)}{810}\,.
\eea
The relation (\ref{betavfromz}) then gives us the five-loop coefficient
of $\be_v=\sum_{k=1}^\infty\de_k g^{k-1}$ through $\de_5=5Z_{15}^v/2$.

\section{\bm Recursion Relations for the $Z_{kl}^v$}
\label{recursion}
The relevant recursion relations among the renormalization constants are
\cite{Ka}
\bea
Z_{25}^v
&=&
\frac{1}{5}[(2Z_{11}^m+3Z_{11}^g)Z_{14}^v
+2(2Z_{12}^m+2Z_{12}^g)Z_{13}^v+3(2Z_{13}^m+Z_{13}^g)Z_{12}^v
+4(2Z_{14}^m)Z_{11}^v]\,,
\\
Z_{35}^v
&=&
\frac{1}{5}[(2Z_{11}^m+3Z_{11}^g)Z_{24}^v
+2(2Z_{12}^m+2Z_{12}^g)Z_{23}^v+3(2Z_{13}^m+Z_{13}^g)Z_{22}^v]\,,
\\
Z_{45}^v
&=&
\frac{1}{5}[(2Z_{11}^m+3Z_{11}^g)Z_{34}^v
+2(2Z_{12}^m+2Z_{12}^g)Z_{33}^v]\,,
\\
Z_{55}^v
&=&
\frac{1}{5}(2Z_{11}^m+3Z_{11}^g)Z_{44}^v\,.
\eea
Using the results from \cite{Ka}, it is straightforward to check that
all of the above relations hold.

\section{\bm$\be_v$ Through Five Loops}
\label{betav}
Combining our result for $\de_5$ with the four-loop result from \cite{Ka},
we arrive at the five-loop result for the $O(N)$-symmetric case,
\bea
\label{betavres}
\be_v(g)
&=&\ts
\frac{N}{4}+\frac{N(N+2)}{96}g^2
+\frac{N(N+2)(N+8)[12\ze(3)-25]}{1296}g^3
\nn\\
&&\ts
+\frac{N(N+2)[-319N^2+13968N+64864
+16(3N^2-382N-1700)\ze(3)+96(4N^2+39N+146)\ze(4)
-1024(5N+22)\ze(5)]}{82944}g^4
\nn\\
&&\ts
+\od(g^5)\,.
\eea

We note that among the beta and gamma functions, $\be_v$ is the easiest
to compute.
Therefore, it would be the prime candidate for the first complete non-trivial
six-loop calculation, since only the last six diagrams in table
\ref{remaininggraphs} have to be computed, of which the first three are
essentially trivial.
After converting the quartically divergent integrals into logarithmically
divergent ones by twice differentiating with respect to $m^2$ (see appendix
\ref{five-loop integrals}), a subset of the diagrams necessary for the
six-loop renormalization of the coupling has to be evaluated.
Since this subset consists of the diagrams where the four external
lines are attached to only two vertices, the most difficult topologies are
absent and the computation should be considerably more easy than the full
coupling renormalization at this level.
A similar statement is true for the comparison with mass and wave function
renormalization.

We further note that the $k$-loop coefficients $\al_k$ of $\ga_m$ we get
as by-products here and in \cite{Ka} come from $\ep^{-1}\ln(m^2/\bar{\mu}^2)$
terms---and therefore originally from $\ep^{-2}$ terms---of $(k+1)$-loop
vacuum diagrams.
Similarly, the $k$-loop coefficients $\be_k$ of $\be_g$ we get come from
$\ep^{-1}\ln^2(m^2/\bar{\mu}^2)$ terms---and therefore originally from
$\ep^{-3}$ terms---of $(k+2)$-loop vacuum diagrams.
These may be easier to compute than the $\ep^{-1}$ terms of
generic $k$-loop diagrams contributing to $\be_k$ and $\al_k$.
In particular, and in contrast to \cite{Sc}, we do not have to consider
the topology of eight-loop vacuum diagrams to obtain the five-loop
contribution to $\be_g$, but only to compute the $1/\ep^3$ terms of
seven-loop vacuum diagrams.
To obtain the five-loop contribution to $\ga_m$, we have to to compute
the $1/\ep^2$ terms of six-loop vacuum diagrams.

However, in the current approach of evaluating loop integrals (to be
described in appendix \ref{five-loop integrals}), the divergences related
to the two-point function at one lower loop (and therefore contributing
to $\ga_m$ at one lower loop) and to the four-point function at two lower
loops  (and therefore contributing to $\be_g$ at two lower loops) than the
considered vacuum graph integrals need to be evaluated as subdivergences of
these vacuum graphs anyway.
Therefore, when evaluating the integrals as described in appendix
\ref{five-loop integrals}, no advantage has been gained for computing
$\ga_m$ and $\be_g$, at least as far as the necessary computational
techniques are concerned.

\section{\bm$\be_v$ and $Z_v$ from the Standard Method}
\label{standard}
Along more traditional lines, we have
\beq\ts
\label{zv}
Z_v=1+\frac{g}{hm^4}{\cal K}\bar{R}\sum\mbox{vacuum graphs}\,,
\eeq
where $\bar{R}$ removes subdivergences of graphs and ${\cal K}$ isolates
the negative powers in $\ep$ (see \cite{CaKe} for a clear definition
of ${\cal K}$ and $\bar{R}$).
The sum in (\ref{zv}) goes over the graphs in table \ref{allgraphs}
and the Feynman rules to be used are
\bea
\bpi(50,12)
\put(5,0){$a$}
\put(39,0){$b$}
\put(15,3){\line(1,0){20}}
\epi
&=&
\frac{\de_{ab}}{p^2+m^2}\,,
\\
\rule[-10pt]{0pt}{28pt}
\bpi(50,12)
\put(5,10){$a$}
\put(39,10){$b$}
\put(5,-10){$c$}
\put(39,-10){$d$}
\put(15,-7){\line(1,1){20}}
\put(15,13){\line(1,-1){20}}
\put(25,3){\circle*{4}}
\epi
&=&
-[\de_{ab}\de_{cd}+\de_{ac}\de_{bd}+\de_{ad}\de_{bc}]
\,\frac{(4\pi)^2 g}{3}
\eea
with renormalized quantities $m^2$ and $g$, since all subdivergences
are removed by the $\bar{R}$ operation.
We have carried out this program from one through five loops in order
to also check the results in \cite{Ka}.
Using the results of appendix \ref{quartdivs}, it is easy to see that
we get the same $Z_v$ and therefore also $\be_v$ as in the approach
above.

If a diagram $D$ separates into $n$ diagrams $D_k$ with independent
integrations, then the standard definition of ${\cal K}\bar{R}$ yields
\beq
\label{sepdivs}
{\cal K}\bar{R}D
={\cal K}\bar{R}\prod_{k=1}^{n}D_k
=(-1)^{n+1}\prod_{k=1}^{n}{\cal K}\bar{R}D_k\,.
\eeq
Since ${\cal K}$ picks out only the pole terms in $\ep$, the $1/\ep$
terms of $Z_v$ and therefore also $\be_v$ receive only contributions
from diagrams whose integrations are not separable.
Thus, for the calculation of $\be_v$, it is again sufficient to
consider the diagrams of table \ref{remaininggraphs}.
The two-loop diagram is no longer needed, though.
However, since, within this program, separable diagrams can be simply put
algebraically together from lower-loop diagrams, including them in the
calculation provides together with the recursion relations between the
$Z_x$ \cite{Ka} a convenient cross-check in the determination of $Z_v$
and $\be_v$.

\section{Cubic Anisotropy}
\label{cubic}
Here we give $\be_v$ for the case of a cubic anisotropy \cite{KlSc,Sc}.
We rename $g\rightarrow g_1$ and introduce a second coupling $g_2$
through
\beq
\label{Lcubic}
{\cal L}
=
\frac{1}{2}\p_\mu\phi_{Bi}\p_\mu\phi_{Bi}
+\frac{1}{2}m_B^2\phi_{Bi}\phi_{Bi}
+\frac{(4\pi)^2}{4!}
\left({g_1}_{\sss\! B}T_{ijkl}^{(1)}+{g_2}_{\sss\! B}T_{ijkl}^{(2)}\right)
\phi_i\phi_j\phi_k\phi_l
+\frac{m_B^4h_B}{(4\pi)^2{g_1}_{\sss\! B}}\,,
\eeq
where repeated indices are summed over
($\mu=1,\ldots,d$ and $i,j,k,l=1,\ldots,N$), the subscript $B$ refers
to bare quantities and where
\bea
\label{t1}
T_{ijkl}^{(1)}
&=&
\frac{1}{3}(\de_{ij}\de_{kl}+\de_{ik}\de_{jl}+\de_{il}\de_{jk})\,,
\\
\label{t2}
T_{ijkl}^{(2)}
&=&
\de_{ijkl}\equiv\left\{\ba{ll}1,&i=j=k=l,\\0,&\mbox{\rm else.}\ea\right.
\eea
Using standard methods \cite{CoMa}, one arrives at
\beq
\label{betav12fromz}
\be_v=\frac{1}{2g_1}\left(
g_1\frac{\p Z_{v,1}}{\p g_1}+g_2\frac{\p Z_{v,1}}{\p g_2}
\right)\,.
\eeq
The seemingly asymmetric treatment of $g_1$ and $g_2$ is entirely due to
our definition of the constant term in (\ref{Lcubic}), since the definitions
(\ref{t1}) and (\ref{t2}) are not used for the derivation of
(\ref{betav12fromz}).
However, it is easy to see that e.g.\ replacing the constant term in
(\ref{Lcubic}) by $m_B^4h_B/[(4\pi)^2{g_2}_{\sss\! B}]$ and redefining
$Z_v$ accordingly exchanges $g_1$ and $g_2$ in (\ref{betav12fromz}), but
leaves $\be_v$ invariant.

Through five loops, we get
\bea
\be_v(g_1,g_2)
&=&\ts
\frac{N}{4}
+\frac{N(N+2)}{96}g_1^2
+\frac{N}{16}g_1g_2
+\frac{N}{32}g_2^2
\nn\\
&&\ts
+\frac{N(N+2)(N+8)[12\ze(3)-25]}{1296}g_1^3
+\frac{N(N+8)[12\ze(3)-25]}{144}g_1^2g_2
+\frac{N[12\ze(3)-25]}{16}g_1g_2^2
+\frac{N[12\ze(3)-25]}{48}g_2^3
\nn\\
&&\ts
+\frac{N(N+2)[-319N^2+13968N+64864
+16(3N^2-382N-1700)\ze(3)+96(4N^2+39N+146)\ze(4)
-1024(5N+22)\ze(5)]}{82944}g_1^4
\nn\\
&&\ts
+\frac{N[-319N^2+13968N+64864+16(3N^2-382N-1700)\ze(3)
+96(4N^2+39N+146)\ze(4)-1024(5N+22)\ze(5)]}{6912}g_1^3g_2
\nn\\
&&\ts
+\frac{N[(6431N+229108)-48(71N+2008)\ze(3)
+864(5N+58)\ze(4)-3072(N+26)\ze(5)]}{13824}g_1^2g_2^2
\nn\\
&&\ts
+\frac{N[26171-11088\ze(3)+6048\ze(4)-9216\ze(5)]}{2304}g_1g_2^3
\nn\\
&&\ts
+\frac{N[26171-11088\ze(3)+6048\ze(4)-9216\ze(5)]}{9216}g_2^4\,,
\eea
which reduces to (\ref{betavres}) for $(g_1,g_2)\rightarrow(g,0)$.

\subsection*{Note Added}
Recently, the result (\ref{betavres}) has been confirmed in an independent
calculation by S.A.~Larin, M.~M\"on\-nig\-mann, M.~Str\"osser and V.~Dohm,
cond-mat/9711069, where also its application to three-dimensional systems
is considered.

\subsection*{Acknowledgements}
I am grateful to Stefan Bornholdt for numerous helpful discussions.
This work was supported by the Deutsche Forschungsgemeinschaft (DFG).

\appendix

\section*{Appendix}

\section{Five-Loop Integrals}
\label{five-loop integrals}
Our strategy for computing the five-loop integrals $I_{5a}$, $I_{5b}$, 
$I_{5c}$ is different from the way of computing integrals in \cite{Ka}.
Here we take two derivatives of each integral with respect to $m^2$ to
convert it into a sum of logarithmically divergent integrals.
Then we subtract subdivergences using the ${\cal K}\bar{R}$ operation,
so that we can set most masses in propagators in the resulting expression
to zero without changing its divergent part.
This allows us to do the relevant integrals using the methods of
infrared rearrangement \cite{Vl}, (our own modified version of) the
$R^*$ operation \cite{ChTkSm} and the results of \cite{Br}, being in turn
partially based also on the integration-by-parts algorithm \cite{ChTk}.
For an introduction to these techniques and the multi-loop renormalization
of $\phi^4$ theory in general, see \cite{Sc}.
At last, we evaluate the terms containing subdivergences that we have
subtracted above and add them again.
These terms have less than five loops and can be computed by either
recursively continuing this procedure or by using the results from
\cite{Ka}.

One might argue that this method of computing integrals is not independent
of the standard method referred to in section \ref{standard}.
Nevertheless, it provides us with many more cross-checks as described
in sections \ref{myscheme} and \ref{recursion}.

All diagrams in this section only refer to momentum space integrals
without symmetry and group factors.
To get the notation in line with appendix \ref{quartdivs}, we rename
the integrals $I_2^{cc}$ and $I_3^{cc}$ from \cite{Ka} $I_{3a}$ and
$I_{4a}$, respectively, so that their finite parts defined in \cite{Ka}
are now $I_{3a,f}=I_{2,f}^{cc}$ and $I_{4a,f}=I_{3,f}^{cc}$.

\subsection{\bm$I_{5a}$}
Define $I_{5a}$ by
\beq
\label{i5adef}
I_{5a}
=
\rule[-18pt]{0pt}{42pt}
\bpi(42,12)
\put(21,3){\circle{32}}
\put(9.7,-8.3){\line(1,0){22.6}}
\put(9.7,-8.3){\line(0,1){22.6}}
\put(9.7,14.3){\line(1,0){22.6}}
\put(32.3,-8.3){\line(0,1){22.6}}
\put(9.7,-8.3){\circle*{4}}
\put(9.7,14.3){\circle*{4}}
\put(32.3,-8.3){\circle*{4}}
\put(32.3,14.3){\circle*{4}}
\epi
=\ts
\Int{kpqrs}\frac{1}{(p^2{+}m^2)[(k{+}p)^2{+}m^2](q^2{+}m^2)
[(k{+}q)^2{+}m^2](r^2{+}m^2)[(k{+}r)^2{+}m^2](s^2{+}m^2)[(k{+}s)^2{+}m^2]}\,.
\eeq
Since $I_{5a}\propto (m^2)^{\frac{5}{2}d-8}
=(m^2)^{2-\frac{5}{2}\ep}$ we can write
\bea
I_{5a}
&=&\ts
\frac{m^4}{(2-\frac{5}{2}\ep)(1-\frac{5}{2}\ep)}
\left(\frac{\p}{\p m^2}\right)^2 I_{5a}
\nn\\
&=&\ts
\frac{8m^4}{(2-\frac{5}{2}\ep)(1-\frac{5}{2}\ep)}
\left[2
\rule[-18pt]{0pt}{42pt}
\bpi(42,12)
\put(21,3){\circle{32}}
\put(9.7,-8.3){\line(1,0){22.6}}
\put(9.7,-8.3){\line(0,1){22.6}}
\put(9.7,14.3){\line(1,0){22.6}}
\put(32.3,-8.3){\line(0,1){22.6}}
\put(9.7,-8.3){\circle*{4}}
\put(9.7,14.3){\circle*{4}}
\put(32.3,-8.3){\circle*{4}}
\put(32.3,14.3){\circle*{4}}
\put(16.9,18.5){\circle*{4}}
\put(25.1,18.5){\circle*{4}}
\epi
+
\rule[-18pt]{0pt}{42pt}
\bpi(42,12)
\put(21,3){\circle{32}}
\put(9.7,-8.3){\line(1,0){22.6}}
\put(9.7,-8.3){\line(0,1){22.6}}
\put(9.7,14.3){\line(1,0){22.6}}
\put(32.3,-8.3){\line(0,1){22.6}}
\put(9.7,-8.3){\circle*{4}}
\put(9.7,14.3){\circle*{4}}
\put(32.3,-8.3){\circle*{4}}
\put(32.3,14.3){\circle*{4}}
\put(21,14.3){\circle*{4}}
\put(21,19){\circle*{4}}
\epi
+6
\rule[-18pt]{0pt}{42pt}
\bpi(42,12)
\put(21,3){\circle{32}}
\put(9.7,-8.3){\line(1,0){22.6}}
\put(9.7,-8.3){\line(0,1){22.6}}
\put(9.7,14.3){\line(1,0){22.6}}
\put(32.3,-8.3){\line(0,1){22.6}}
\put(9.7,-8.3){\circle*{4}}
\put(9.7,14.3){\circle*{4}}
\put(32.3,-8.3){\circle*{4}}
\put(32.3,14.3){\circle*{4}}
\put(21,-13){\circle*{4}}
\put(21,19){\circle*{4}}
\epi
\right]\,.
\eea
The three integrals on the right hand side can now be evaluated as
described above.
The result is
\bea
\label{i5ares}
I_{5a}
&=&\ts
\frac{m^4}{(4\pi)^{10}}
\left\{
\frac{192}{5\ep^5}
+\frac{1}{\ep^4}\left[-96\lnma+\frac{848}{5}\right]
+\frac{1}{\ep^3}\left[120\lnmb-424\lnma
+\left(\frac{2004}{5}+24\ze(2)\right)\right]\right.
\nn\\
&&\ts\hspace{35pt}
+\frac{1}{\ep^2}\left[116\lnmc-298\lnmb+(258+156\ze(2))\lnma
+\left(-\frac{317}{5}-170\ze(2)+\frac{88}{5}\ze(3)\right)\right]
\nn\\
&&\ts
\hspace{35pt}+\frac{1}{\ep}\left[
\frac{31}{2}\lnmd-45\lnmc+\left(\frac{107}{2}+45\ze(2)\right)\lnmb
-\left(\frac{19}{2}+71\ze(2)+12\ze(3)\right)\lnma
\right.
\nn\\
&&\ts
\hspace{55pt}\left.\left.
+\left(-\frac{609}{20}+\frac{59}{2}\ze(2)+\frac{82}{5}\ze(3)
+\frac{141}{5}\ze(4)+\frac{21}{2}\ze(2)^2\right)\right]\right\}
\nn\\
&&\ts
+\frac{24}{(4\pi)^4\ep^2}I_{3a,f}
+\frac{8}{(4\pi)^2\ep}I_{4a,f}+I_{5a,f}\,,
\eea
where $I_{5a,f}=\od(\ep^0)$.

\subsection{\bm$I_{5b}$}
Define $I_{5b}$ by
\beq
\label{i5bdef}
I_{5b}
=
\rule[-22pt]{0pt}{50pt}
\bpi(42,12)
\put(21,-9){\circle{16}}
\put(21,15){\circle{16}}
\put(13,-9){\line(1,0){16}}
\put(13,15){\line(1,0){16}}
\put(13,3){\oval(16,24)[l]}
\put(29,3){\oval(16,24)[r]}
\put(13,-9){\circle*{4}}
\put(13,15){\circle*{4}}
\put(29,-9){\circle*{4}}
\put(29,15){\circle*{4}}
\epi
=\ts
\Int{kpqrs}\frac{1}{(k^2{+}m^2)^2
[(k{+}p{+}q)^2{+}m^2](p^2{+}m^2)(q^2{+}m^2)
[(k{+}r{+}s)^2{+}m^2](r^2{+}m^2)(s^2{+}m^2)}\,.
\eeq
We can write
\bea
\label{i5b}
I_{5b}
&=&\ts
\frac{m^4}{(2-\frac{5}{2}\ep)(1-\frac{5}{2}\ep)}
\left(\frac{\p}{\p m^2}\right)^2 I_{5b}
\nn\\
&=&\ts
\frac{6m^4}{(2-\frac{5}{2}\ep)(1-\frac{5}{2}\ep)}
\left[2
\rule[-22pt]{0pt}{50pt}
\bpi(42,12)
\put(21,-9){\circle{16}}
\put(21,15){\circle{16}}
\put(13,-9){\line(1,0){16}}
\put(13,15){\line(1,0){16}}
\put(13,3){\oval(16,24)[l]}
\put(29,3){\oval(16,24)[r]}
\put(13,-9){\circle*{4}}
\put(13,15){\circle*{4}}
\put(29,-9){\circle*{4}}
\put(29,15){\circle*{4}}
\put(17,21.9){\circle*{4}}
\put(25,21.9){\circle*{4}}
\epi
+2
\rule[-22pt]{0pt}{50pt}
\bpi(42,12)
\put(21,-9){\circle{16}}
\put(21,15){\circle{16}}
\put(13,-9){\line(1,0){16}}
\put(13,15){\line(1,0){16}}
\put(13,3){\oval(16,24)[l]}
\put(29,3){\oval(16,24)[r]}
\put(13,-9){\circle*{4}}
\put(13,15){\circle*{4}}
\put(29,-9){\circle*{4}}
\put(29,15){\circle*{4}}
\put(21,23){\circle*{4}}
\put(21,7){\circle*{4}}
\epi
+3
\rule[-22pt]{0pt}{50pt}
\bpi(42,12)
\put(21,-9){\circle{16}}
\put(21,15){\circle{16}}
\put(13,-9){\line(1,0){16}}
\put(13,15){\line(1,0){16}}
\put(13,3){\oval(16,24)[l]}
\put(29,3){\oval(16,24)[r]}
\put(13,-9){\circle*{4}}
\put(13,15){\circle*{4}}
\put(29,-9){\circle*{4}}
\put(29,15){\circle*{4}}
\put(21,-17){\circle*{4}}
\put(21,23){\circle*{4}}
\epi
+4
\rule[-22pt]{0pt}{50pt}
\bpi(42,12)
\put(21,-9){\circle{16}}
\put(21,15){\circle{16}}
\put(13,-9){\line(1,0){16}}
\put(13,15){\line(1,0){16}}
\put(13,3){\oval(16,24)[l]}
\put(29,3){\oval(16,24)[r]}
\put(13,-9){\circle*{4}}
\put(13,15){\circle*{4}}
\put(29,-9){\circle*{4}}
\put(29,15){\circle*{4}}
\put(5,3){\circle*{4}}
\put(21,23){\circle*{4}}
\epi
+
\rule[-22pt]{0pt}{50pt}
\bpi(42,12)
\put(21,-9){\circle{16}}
\put(21,15){\circle{16}}
\put(13,-9){\line(1,0){16}}
\put(13,15){\line(1,0){16}}
\put(13,3){\oval(16,24)[l]}
\put(29,3){\oval(16,24)[r]}
\put(13,-9){\circle*{4}}
\put(13,15){\circle*{4}}
\put(29,-9){\circle*{4}}
\put(29,15){\circle*{4}}
\put(5,3){\circle*{4}}
\put(37,3){\circle*{4}}
\epi
\right]
\eea
and evaluate this to give
\bea
\label{i5bres}
I_{5b}
&=&\ts
\frac{m^4}{(4\pi)^{10}}
\left\{
\frac{192}{5\ep^5}
+\frac{1}{\ep^4}\left[-96\lnma+\frac{452}{5}\right]
+\frac{1}{\ep^3}\left[120\lnmb-226\lnma
+\left(\frac{521}{5}+24\ze(2)\right)\right]\right.
\nn\\
&&\ts\hspace{30pt}
+\frac{1}{\ep^2}\left[-46\lnmc+\frac{151}{2}\lnmb
+\left(\frac{109}{2}-6\ze(2)\right)\lnma
+\left(-\frac{2463}{20}-\frac{25}{2}\ze(2)+4\ze(3)\right)\right]
\nn\\
&&\ts\hspace{30pt}
+\frac{1}{\ep}\left[
-\frac{47}{4}\lnmd+\frac{1225}{12}\lnmc
-\left(\frac{2633}{8}+\frac{39}{2}\ze(2)\right)\lnmb
+\left(\frac{931}{2}+\frac{233}{4}\ze(2)-10\ze(3)\right)\lnma
\right.
\nn\\
&&\ts\hspace{50pt}
\left.\left.
+\left(-\frac{1214}{5}-\frac{385}{8}\ze(2)+\frac{61}{6}\ze(3)
-\frac{3}{2}\ze(4)+\frac{3}{4}\ze(2)^2\right)
\rule{0pt}{17pt}\right]\right\}
\nn\\
&&\ts
+\frac{6}{(4\pi)^4}
\left[\frac{1}{\ep^2}+\frac{1}{\ep}\left(-\lnma+\frac{1}{2}\right)\right]
I_{3a,f}+I_{5b,f}\,,
\eea
where $I_{5b,f}=\od(\ep^0)$.

\subsection{\bm$I_{5c}$}
Define $I_{5c}$ by
\beq
\label{i5cdef}
I_{5c}
\equiv
\rule[-18pt]{0pt}{42pt}
\bpi(58,12)
\put(13,3){\circle{16}}
\put(45,3){\circle{16}}
\put(5,-5){\line(0,1){16}}
\put(25,-5){\oval(40,16)[b]}
\put(25,11){\oval(40,16)[t]}
\put(45,3){\oval(48,16)[l]}
\put(5,3){\circle*{4}}
\put(21,3){\circle*{4}}
\put(45,-5){\circle*{4}}
\put(45,11){\circle*{4}}
\epi
=\ts
\Int{kpqrs}\frac{1}{
(p^2+m^2)[(p+k)^2+m^2](q^2+m^2)[(q+k)^2+m^2]
(r^2+m^2)[(r+p-q)^2+m^2](s^2+m^2)[(s+k)^2+m^2]}\,.
\eeq
We can write
\bea
\label{i5c}
I_{5c}
&=&\ts
\frac{m^4}{(2-\frac{5}{2}\ep)(1-\frac{5}{2}\ep)}
\left(\frac{\p}{\p m^2}\right)^2 I_{5c}
\nn\\
&=&\ts
\frac{4m^4}{(2-\frac{5}{2}\ep)(1-\frac{5}{2}\ep)}
\left[2
\rule[-18pt]{0pt}{42pt}
\bpi(58,12)
\put(13,3){\circle{16}}
\put(45,3){\circle{16}}
\put(5,-5){\line(0,1){16}}
\put(25,-5){\oval(40,16)[b]}
\put(25,11){\oval(40,16)[t]}
\put(45,3){\oval(48,16)[l]}
\put(5,3){\circle*{4}}
\put(21,3){\circle*{4}}
\put(45,-5){\circle*{4}}
\put(45,11){\circle*{4}}
\put(51.9,-1){\circle*{4}}
\put(51.9,7){\circle*{4}}
\epi
+
\rule[-18pt]{0pt}{42pt}
\bpi(58,12)
\put(13,3){\circle{16}}
\put(45,3){\circle{16}}
\put(5,-5){\line(0,1){16}}
\put(25,-5){\oval(40,16)[b]}
\put(25,11){\oval(40,16)[t]}
\put(45,3){\oval(48,16)[l]}
\put(5,3){\circle*{4}}
\put(21,3){\circle*{4}}
\put(45,-5){\circle*{4}}
\put(45,11){\circle*{4}}
\put(37,3){\circle*{4}}
\put(53,3){\circle*{4}}
\epi
+8
\rule[-18pt]{0pt}{42pt}
\bpi(58,12)
\put(13,3){\circle{16}}
\put(45,3){\circle{16}}
\put(5,-5){\line(0,1){16}}
\put(25,-5){\oval(40,16)[b]}
\put(25,11){\oval(40,16)[t]}
\put(45,3){\oval(48,16)[l]}
\put(5,3){\circle*{4}}
\put(21,3){\circle*{4}}
\put(45,-5){\circle*{4}}
\put(45,11){\circle*{4}}
\put(53,3){\circle*{4}}
\put(25,19){\circle*{4}}
\epi
+2
\rule[-18pt]{0pt}{42pt}
\bpi(58,12)
\put(13,3){\circle{16}}
\put(45,3){\circle{16}}
\put(5,-5){\line(0,1){16}}
\put(25,-5){\oval(40,16)[b]}
\put(25,11){\oval(40,16)[t]}
\put(45,3){\oval(48,16)[l]}
\put(5,3){\circle*{4}}
\put(21,3){\circle*{4}}
\put(45,-5){\circle*{4}}
\put(45,11){\circle*{4}}
\put(13,11){\circle*{4}}
\put(53,3){\circle*{4}}
\epi
+2
\rule[-18pt]{0pt}{42pt}
\bpi(58,12)
\put(13,3){\circle{16}}
\put(45,3){\circle{16}}
\put(5,-5){\line(0,1){16}}
\put(25,-5){\oval(40,16)[b]}
\put(25,11){\oval(40,16)[t]}
\put(45,3){\oval(48,16)[l]}
\put(5,3){\circle*{4}}
\put(21,3){\circle*{4}}
\put(45,-5){\circle*{4}}
\put(45,11){\circle*{4}}
\put(33,11){\circle*{4}}
\put(25,19){\circle*{4}}
\epi
\right.
\nn\\
&&\ts\hspace{60pt}
\left.+
\rule[-18pt]{0pt}{42pt}
\bpi(58,12)
\put(13,3){\circle{16}}
\put(45,3){\circle{16}}
\put(5,-5){\line(0,1){16}}
\put(25,-5){\oval(40,16)[b]}
\put(25,11){\oval(40,16)[t]}
\put(45,3){\oval(48,16)[l]}
\put(5,3){\circle*{4}}
\put(21,3){\circle*{4}}
\put(45,-5){\circle*{4}}
\put(45,11){\circle*{4}}
\put(33,-5){\circle*{4}}
\put(25,19){\circle*{4}}
\epi
+2
\rule[-18pt]{0pt}{42pt}
\bpi(58,12)
\put(13,3){\circle{16}}
\put(45,3){\circle{16}}
\put(5,-5){\line(0,1){16}}
\put(25,-5){\oval(40,16)[b]}
\put(25,11){\oval(40,16)[t]}
\put(45,3){\oval(48,16)[l]}
\put(5,3){\circle*{4}}
\put(21,3){\circle*{4}}
\put(45,-5){\circle*{4}}
\put(45,11){\circle*{4}}
\put(18.3,19){\circle*{4}}
\put(31.7,19){\circle*{4}}
\epi
\right]
\eea
and evaluate this to give
\bea
\label{i5cres}
I_{5c}
&=&\ts
\frac{m^4}{(4\pi)^{10}}
\bigg\{
\frac{96}{5\ep^5}
+\frac{1}{\ep^4}\left[-48\lnma+\frac{456}{5}\right]
+\frac{1}{\ep^3}\left[60\lnmb-228\lnma
+\frac{3542}{15}+12\ze(2)+\frac{32}{5}\ze(3)\right]
\nn\\
&&\ts\hspace{30pt}
+\frac{1}{\ep^2}\bigg[58\lnmc-129\lnmb
+\left(\frac{119}{3}+78\ze(2)-16\ze(3)\right)\lnma
\nn\\
&&\ts\hspace{55pt}
+\left(\frac{457}{6}-81\ze(2)+\frac{84}{5}\ze(3)
+\frac{24}{5}\ze(4)\right)\bigg]
\nn\\
&&\ts\hspace{30pt}
+\frac{1}{\ep}\bigg[
\frac{31}{4}\lnmd-\frac{19}{6}\lnmc
+\left(\frac{5}{12}+\frac{45}{2}\ze(2)+20\ze(3)\right)\lnmb
\nn\\
&&\ts\hspace{55pt}
+\left(-\frac{773}{12}-\frac{19}{2}\ze(2)-26\ze(3)-12\ze(4)\right)\lnma
\nn\\
&&\ts\hspace{55pt}
+\left(\frac{2309}{24}-\frac{107}{12}\ze(2)+\frac{5}{3}\ze(3)
+\frac{201}{10}\ze(4)-\frac{64}{5}\ze(5)+\frac{21}{4}\ze(2)^2
+4\ze(2)\ze(3)\right)\bigg]\bigg\}
\nn\\
&&\ts
+\frac{1}{(4\pi)^4}
\left(\frac{12}{\ep^2}+\frac{4}{\ep}\right)I_{3a,f}
+\frac{4}{(4\pi)^2\ep}I_{4a,f}+I_{5c,f}\,,
\eea
where $I_{5c,f}=\od(\ep^0)$.

\section{\bm${\cal K}\bar{R}$ on Quartically Divergent Diagrams}
\label{quartdivs}
Using standard methods, we have evaluated the quartically divergent
diagrams from table \ref{allgraphs} with subdivergences removed,
as needed for the determination of $Z_v$ in section \ref{standard}. 
As in the last section, the diagrams in the following refer to momentum
space integrals, while symmetry factors $S_x$ and group factors
$G_x(g_1,g_2)$ are written out separately.
As indicated, we define the group factors to contain not only the
contributions from the contraction of the $\de_{ij}$ and $\de_{ijkl}$
tensors, but also their accompanying factors $-g_1/3$ and $-g_2$,
respectively.
For diagrams with separable integrations, we only give the
$O(N)$-symmetric result $G_x(g,0)$, since these diagrams are exclusively
used for cross-checks, which we have performed only for the
$O(N)$-symmetric case.
For diagrams whose integrations do not separate, we give $G_x(g_1,g_2)$.

Our strategy to compute the quartically divergent integrals with
subdivergences removed is to convert quartically divergent integrals by
twice differentiating with respect to $m^2$ to logarithmically divergent
ones and subsequently evaluate these by the methods of
\cite{Vl}-\cite{ChTk}.
We exploit the fact that the operator $\p/\p m^2$ commutes with
${\cal K}\bar{R}$ (in the same way as differentiation with respect
to an external momentum commutes with ${\cal K}\bar{R}$, see \cite{CaKe})
and that the result of ${\cal K}\bar{R}$ acting on a quartically divergent
diagram is proportional to $m^4$,
\beq
{\cal K}\bar{R}
\rule[-14pt]{0pt}{34pt}
\bpi(34,12)
\put(17,3){\circle{24}}
\put(5.02,2.33){\line(1,1){12.65}}
\put(6.17,-2.17){\line(1,1){16}}
\put(8.51,-5.49){\line(1,1){16.97}}
\put(11.83,-7.83){\line(1,1){16}}
\put(16.33,-8.98){\line(1,1){12.65}}
\epi
=
\frac{1}{2}m^4{\cal K}\bar{R}\left(\frac{\p}{\p m^2}\right)^2
\rule[-14pt]{0pt}{34pt}
\bpi(34,12)
\put(17,3){\circle{24}}
\put(5.02,2.33){\line(1,1){12.65}}
\put(6.17,-2.17){\line(1,1){16}}
\put(8.51,-5.49){\line(1,1){16.97}}
\put(11.83,-7.83){\line(1,1){16}}
\put(16.33,-8.98){\line(1,1){12.65}}
\epi\,.
\eeq
We have evaluated all logarithmically divergent diagrams encountered
en route and found agreement with the respective results in \cite{Sc}.
Therefore, we do not display the results for the logarithmically
divergent diagrams here.
In order to also check our results in \cite{Ka}, we have carried out
this program from one through five loops and give the results for
all diagrams of table \ref{allgraphs} below.

The notation of $\bar{I}_x$ refers to the integral $I_x$ with 
subdivergences removed by ${\cal K}\bar{R}$.
The integrals $I_{3a}$ and $I_{4a}$ below are identical to
$I_2^{cc}$ and $I_3^{cc}$ in \cite{Ka}, respectively.

\subsection{One Loop}
\beq\ts
\bar{I}_{1a}
\equiv
{\cal K}\bar{R}
\rule[-10pt]{0pt}{26pt}
\bpi(26,12)
\put(13,3){\circle{16}}
\epi
=
\frac{1}{2}m^4
{\cal K}\bar{R}
\rule[-10pt]{0pt}{26pt}
\bpi(26,12)
\put(13,3){\circle{16}}
\put(5,3){\circle*{4}}
\put(21,3){\circle*{4}}
\epi
\nn\\
=
\frac{m^4}{(4\pi)^2\ep}\,,
\eeq

\beq\ts
S_{1a}=\frac{1}{2}\,,
\eeq
\beq
G_{1a}(g_1,g_2)=N\,.
\eeq

\subsection{Two Loops}
\beq\ts
\bar{I}_{2a}
\equiv
{\cal K}\bar{R}
\rule[-10pt]{0pt}{26pt}
\bpi(42,12)
\put(13,3){\circle{16}}
\put(29,3){\circle{16}}
\put(21,3){\circle*{4}}
\epi
=
m^4{\cal K}\bar{R}
\rule[-10pt]{0pt}{26pt}
\bpi(42,12)
\put(13,3){\circle{16}}
\put(29,3){\circle{16}}
\put(5,3){\circle*{4}}
\put(21,3){\circle*{4}}
\put(37,3){\circle*{4}}
\epi
\nn\\
=
-\frac{4m^4}{(4\pi)^4\ep^2}\,,
\eeq

\beq\ts
S_{2a}=\frac{1}{8}\,,
\eeq
\beq\ts
G_{2a}(g,0)=-\frac{1}{3}N(N+2)g\,.
\eeq

\subsection{Three Loops}
\beq\ts
\bar{I}_{3a}
\equiv
{\cal K}\bar{R}
\rule[-14pt]{0pt}{34pt}
\bpi(34,12)
\put(17,3){\circle{24}}
\put(17,3){\oval(24,8)}
\put(5,3){\circle*{4}}
\put(29,3){\circle*{4}}
\epi
=
\frac{1}{2}m^4{\cal K}\bar{R}
\left[
8\rule[-14pt]{0pt}{34pt}
\bpi(34,12)
\put(17,3){\circle{24}}
\put(17,3){\oval(24,8)}
\put(5,3){\circle*{4}}
\put(29,3){\circle*{4}}
\put(11,13.4){\circle*{4}}
\put(23,13.4){\circle*{4}}
\epi
+
12\rule[-14pt]{0pt}{34pt}
\bpi(34,12)
\put(17,3){\circle{24}}
\put(17,3){\oval(24,8)}
\put(5,3){\circle*{4}}
\put(17,-9){\circle*{4}}
\put(17,15){\circle*{4}}
\put(29,3){\circle*{4}}
\epi
\right]
\nn\\
=
\frac{m^4}{(4\pi)^6}
\left(\frac{16}{\ep^3}-\frac{40}{3\ep^2}+\frac{1}{\ep}\right)\,,
\eeq

\beq\ts
S_{3a}=\frac{1}{48}\,,
\eeq
\beq\ts
G_{3a}(g_1,g_2)=\frac{1}{3}N(N+2)g_1^2+2Ng_1g_2+Ng_2^2\,.
\eeq

\beq\ts
\bar{I}_{3b}
\equiv
{\cal K}\bar{R}
\rule[-10pt]{0pt}{26pt}
\bpi(58,12)
\put(13,3){\circle{16}}
\put(29,3){\circle{16}}
\put(45,3){\circle{16}}
\put(21,3){\circle*{4}}
\put(37,3){\circle*{4}}
\epi
=
m^4{\cal K}\bar{R}
\rule[-10pt]{0pt}{26pt}
\bpi(58,12)
\put(13,3){\circle{16}}
\put(29,3){\circle{16}}
\put(45,3){\circle{16}}
\put(5,3){\circle*{4}}
\put(21,3){\circle*{4}}
\put(37,3){\circle*{4}}
\put(53,3){\circle*{4}}
\epi
\nn\\
=
\frac{8m^4}{(4\pi)^6\ep^3}\,,
\eeq

\beq\ts
S_{3b}=\frac{1}{16}\,,
\eeq
\beq\ts
G_{3b}(g,0)=\frac{1}{9}N(N+2)^2g^2\,.
\eeq

\subsection{Four Loops}
\bea
\bar{I}_{4a}
&\equiv&\ts
{\cal K}\bar{R}
\rule[-14pt]{0pt}{34pt}
\bpi(34,12)
\put(17,3){\circle{24}}
\put(6.6,9){\line(1,0){20.8}}
\put(6.6,9){\line(3,-5){10.4}}
\put(27.4,9){\line(-3,-5){10.4}}
\put(6.6,9){\circle*{4}}
\put(27.4,9){\circle*{4}}
\put(17,-9){\circle*{4}}
\epi
=
\frac{1}{2}m^4{\cal K}\bar{R}
\left[
12\rule[-14pt]{0pt}{34pt}
\bpi(34,12)
\put(17,3){\circle{24}}
\put(6.6,9){\line(1,0){20.8}}
\put(6.6,9){\line(3,-5){10.4}}
\put(27.4,9){\line(-3,-5){10.4}}
\put(6.6,9){\circle*{4}}
\put(27.4,9){\circle*{4}}
\put(17,-9){\circle*{4}}
\put(12.9,14.3){\circle*{4}}
\put(21.3,14.3){\circle*{4}}
\epi
+
6\rule[-14pt]{0pt}{34pt}
\bpi(34,12)
\put(17,3){\circle{24}}
\put(6.6,9){\line(1,0){20.8}}
\put(6.6,9){\line(3,-5){10.4}}
\put(27.4,9){\line(-3,-5){10.4}}
\put(6.6,9){\circle*{4}}
\put(27.4,9){\circle*{4}}
\put(17,-9){\circle*{4}}
\put(17,9){\circle*{4}}
\put(17,15){\circle*{4}}
\epi
+
24\rule[-14pt]{0pt}{34pt}
\bpi(34,12)
\put(17,3){\circle{24}}
\put(6.6,9){\line(1,0){20.8}}
\put(6.6,9){\line(3,-5){10.4}}
\put(27.4,9){\line(-3,-5){10.4}}
\put(6.6,9){\circle*{4}}
\put(27.4,9){\circle*{4}}
\put(17,-9){\circle*{4}}
\put(6.6,-3){\circle*{4}}
\put(27.4,-3){\circle*{4}}
\epi
\right]
\nn\\
&=&\ts
\frac{m^4}{(4\pi)^8}
\left[-\frac{24}{\ep^4}+\frac{44}{\ep^3}-\frac{42}{\ep^2}
+\frac{1}{\ep}\left(\frac{25}{2}-6\ze(3)\right)\right]\,,
\eea

\beq\ts
S_{4a}=\frac{1}{48}\,,
\eeq
\beq\ts
G_{4a}(g_1,g_2)=-\frac{1}{27}N(N+2)(N+8)g_1^3-\frac{1}{3}N(N+8)g_1^2g_2
-3Ng_1g_2^2-Ng_2^3\,.
\eeq

\beq\ts
\bar{I}_{4b}
\equiv
{\cal K}\bar{R}
\rule[-14pt]{0pt}{50pt}
\bpi(34,12)
\put(17,3){\circle{24}}
\put(17,3){\oval(24,8)}
\put(17,23){\circle{16}}
\put(5,3){\circle*{4}}
\put(29,3){\circle*{4}}
\put(17,15){\circle*{4}}
\epi
=
\frac{1}{2}m^4{\cal K}\bar{R}
\left[
4\rule[-14pt]{0pt}{47.3pt}
\bpi(46,12)
\put(23,3){\circle{24}}
\put(13,20.3){\circle{16}}
\put(23,3){\oval(24,8)}
\put(11,3){\circle*{4}}
\put(35,3){\circle*{4}}
\put(17,13.4){\circle*{4}}
\put(29,13.4){\circle*{4}}
\put(9,27.25){\circle*{4}}
\epi
+
6\rule[-14pt]{0pt}{50pt}
\bpi(34,12)
\put(17,3){\circle{24}}
\put(17,3){\oval(24,8)}
\put(17,23){\circle{16}}
\put(5,3){\circle*{4}}
\put(17,15){\circle*{4}}
\put(17,31){\circle*{4}}
\put(29,3){\circle*{4}}
\put(17,-9){\circle*{4}}
\epi
\right]
=
\frac{m^4}{(4\pi)^8}
\left(-\frac{16}{\ep^4}+\frac{40}{3\ep^3}-\frac{1}{\ep^2}\right)\,,
\eeq

\beq\ts
S_{4b}=\frac{1}{24}\,,
\eeq
\beq\ts
G_{4b}(g,0)=-\frac{1}{9}N(N+2)^2g^3\,.
\eeq

\beq\ts
\bar{I}_{4c}
\equiv
{\cal K}\bar{R}
\rule[-10pt]{0pt}{26pt}
\bpi(74,12)
\put(13,3){\circle{16}}
\put(29,3){\circle{16}}
\put(45,3){\circle{16}}
\put(61,3){\circle{16}}
\put(21,3){\circle*{4}}
\put(37,3){\circle*{4}}
\put(53,3){\circle*{4}}
\epi
=
m^4{\cal K}\bar{R}
\rule[-10pt]{0pt}{26pt}
\bpi(74,12)
\put(13,3){\circle{16}}
\put(29,3){\circle{16}}
\put(45,3){\circle{16}}
\put(61,3){\circle{16}}
\put(5,3){\circle*{4}}
\put(21,3){\circle*{4}}
\put(37,3){\circle*{4}}
\put(53,3){\circle*{4}}
\put(69,3){\circle*{4}}
\epi
=
-\frac{16m^4}{(4\pi)^8\ep^4}\,,
\eeq

\beq\ts
S_{4c}=\frac{1}{32}\,,
\eeq
\beq\ts
G_{4c}(g,0)=-\frac{1}{27}N(N+2)^3g^3\,.
\eeq

\beq\ts
\bar{I}_{4d}
\equiv
{\cal K}\bar{R}
\rule[-18pt]{0pt}{50pt}
\bpi(53.7,12)
\put(26.85,3){\circle{16}}
\put(26.85,19){\circle{16}}
\put(13,-5){\circle{16}}
\put(40.7,-5){\circle{16}}
\put(26.85,11){\circle*{4}}
\put(19.95,-1){\circle*{4}}
\put(33.75,-1){\circle*{4}}
\epi
=
0\,,
\eeq

\beq\ts
S_{4d}=\frac{1}{48}
\eeq
\beq\ts
G_{4d}(g,0)=-\frac{1}{27}N(N+2)^3g^3,.
\eeq

\subsection{Five Loops}
\bea
\bar{I}_{5a}
&\equiv&\ts
{\cal K}\bar{R}
\rule[-18pt]{0pt}{42pt}
\bpi(42,12)
\put(21,3){\circle{32}}
\put(9.7,-8.3){\line(1,0){22.6}}
\put(9.7,-8.3){\line(0,1){22.6}}
\put(9.7,14.3){\line(1,0){22.6}}
\put(32.3,-8.3){\line(0,1){22.6}}
\put(9.7,-8.3){\circle*{4}}
\put(9.7,14.3){\circle*{4}}
\put(32.3,-8.3){\circle*{4}}
\put(32.3,14.3){\circle*{4}}
\epi
=
\frac{1}{2}m^4{\cal K}\bar{R}\left[
16\rule[-18pt]{0pt}{42pt}
\bpi(42,12)
\put(21,3){\circle{32}}
\put(9.7,-8.3){\line(1,0){22.6}}
\put(9.7,-8.3){\line(0,1){22.6}}
\put(9.7,14.3){\line(1,0){22.6}}
\put(32.3,-8.3){\line(0,1){22.6}}
\put(9.7,-8.3){\circle*{4}}
\put(9.7,14.3){\circle*{4}}
\put(32.3,-8.3){\circle*{4}}
\put(32.3,14.3){\circle*{4}}
\put(16.9,18.5){\circle*{4}}
\put(25.1,18.5){\circle*{4}}
\epi
+
8\rule[-18pt]{0pt}{42pt}
\bpi(42,12)
\put(21,3){\circle{32}}
\put(9.7,-8.3){\line(1,0){22.6}}
\put(9.7,-8.3){\line(0,1){22.6}}
\put(9.7,14.3){\line(1,0){22.6}}
\put(32.3,-8.3){\line(0,1){22.6}}
\put(9.7,-8.3){\circle*{4}}
\put(9.7,14.3){\circle*{4}}
\put(32.3,-8.3){\circle*{4}}
\put(32.3,14.3){\circle*{4}}
\put(21,14.3){\circle*{4}}
\put(21,19){\circle*{4}}
\epi
+
48\rule[-18pt]{0pt}{42pt}
\bpi(42,12)
\put(21,3){\circle{32}}
\put(9.7,-8.3){\line(1,0){22.6}}
\put(9.7,-8.3){\line(0,1){22.6}}
\put(9.7,14.3){\line(1,0){22.6}}
\put(32.3,-8.3){\line(0,1){22.6}}
\put(9.7,-8.3){\circle*{4}}
\put(9.7,14.3){\circle*{4}}
\put(32.3,-8.3){\circle*{4}}
\put(32.3,14.3){\circle*{4}}
\put(21,-13){\circle*{4}}
\put(21,19){\circle*{4}}
\epi
\right]
\nn\\
&=&\ts
\frac{m^4}{(4\pi)^{10}}
\left[\frac{192}{5\ep^5}-\frac{352}{5\ep^4}+\frac{144}{5\ep^3}
+\frac{1}{\ep^2}\left(\frac{168}{5}-\frac{272}{5}\ze(3)\right)
+\frac{1}{\ep}\left(-\frac{319}{20}+\frac{12}{5}\ze(3)
+\frac{96}{5}\ze(4)\right)\right]\,,
\nn\\
\eea

\beq\ts 
S_{5a}=\frac{1}{128}\,,
\eeq
\beq\ts
G_{5a}(g_1,g_2)=\frac{1}{81}N(N+2)(N^2+6N+20)g_1^4
+\frac{4}{27}N(N^2+6N+20)g_1^3g_2
+\frac{2}{3}N(N+8)g_1^2g_2^2+4Ng_1g_2^3+Ng_2^4\,.
\eeq

\bea
\bar{I}_{5b}
&\equiv&\ts
{\cal K}\bar{R}
\rule[-22pt]{0pt}{50pt}
\bpi(42,12)
\put(21,-9){\circle{16}}
\put(21,15){\circle{16}}
\put(13,-9){\line(1,0){16}}
\put(13,15){\line(1,0){16}}
\put(13,3){\oval(16,24)[l]}
\put(29,3){\oval(16,24)[r]}
\put(13,-9){\circle*{4}}
\put(13,15){\circle*{4}}
\put(29,-9){\circle*{4}}
\put(29,15){\circle*{4}}
\epi
=
\frac{1}{2}m^4{\cal K}\bar{R}\left[
12\rule[-22pt]{0pt}{50pt}
\bpi(42,12)
\put(21,-9){\circle{16}}
\put(21,15){\circle{16}}
\put(13,-9){\line(1,0){16}}
\put(13,15){\line(1,0){16}}
\put(13,3){\oval(16,24)[l]}
\put(29,3){\oval(16,24)[r]}
\put(13,-9){\circle*{4}}
\put(13,15){\circle*{4}}
\put(29,-9){\circle*{4}}
\put(29,15){\circle*{4}}
\put(17,21.9){\circle*{4}}
\put(25,21.9){\circle*{4}}
\epi
+
12\rule[-22pt]{0pt}{50pt}
\bpi(42,12)
\put(21,-9){\circle{16}}
\put(21,15){\circle{16}}
\put(13,-9){\line(1,0){16}}
\put(13,15){\line(1,0){16}}
\put(13,3){\oval(16,24)[l]}
\put(29,3){\oval(16,24)[r]}
\put(13,-9){\circle*{4}}
\put(13,15){\circle*{4}}
\put(29,-9){\circle*{4}}
\put(29,15){\circle*{4}}
\put(21,23){\circle*{4}}
\put(21,7){\circle*{4}}
\epi
+
18\rule[-22pt]{0pt}{50pt}
\bpi(42,12)
\put(21,-9){\circle{16}}
\put(21,15){\circle{16}}
\put(13,-9){\line(1,0){16}}
\put(13,15){\line(1,0){16}}
\put(13,3){\oval(16,24)[l]}
\put(29,3){\oval(16,24)[r]}
\put(13,-9){\circle*{4}}
\put(13,15){\circle*{4}}
\put(29,-9){\circle*{4}}
\put(29,15){\circle*{4}}
\put(21,-17){\circle*{4}}
\put(21,23){\circle*{4}}
\epi
+
24\rule[-22pt]{0pt}{50pt}
\bpi(42,12)
\put(21,-9){\circle{16}}
\put(21,15){\circle{16}}
\put(13,-9){\line(1,0){16}}
\put(13,15){\line(1,0){16}}
\put(13,3){\oval(16,24)[l]}
\put(29,3){\oval(16,24)[r]}
\put(13,-9){\circle*{4}}
\put(13,15){\circle*{4}}
\put(29,-9){\circle*{4}}
\put(29,15){\circle*{4}}
\put(5,3){\circle*{4}}
\put(21,23){\circle*{4}}
\epi
+
6\rule[-22pt]{0pt}{50pt}
\bpi(42,12)
\put(21,-9){\circle{16}}
\put(21,15){\circle{16}}
\put(13,-9){\line(1,0){16}}
\put(13,15){\line(1,0){16}}
\put(13,3){\oval(16,24)[l]}
\put(29,3){\oval(16,24)[r]}
\put(13,-9){\circle*{4}}
\put(13,15){\circle*{4}}
\put(29,-9){\circle*{4}}
\put(29,15){\circle*{4}}
\put(5,3){\circle*{4}}
\put(37,3){\circle*{4}}
\epi
\right]
\nn\\
&=&\ts
\frac{m^4}{(4\pi)^{10}}
\left(\frac{192}{5\ep^5}-\frac{208}{5\ep^4}-\frac{74}{5\ep^3}
+\frac{138}{5\ep^2}-\frac{71}{20\ep}\right)\,,
\eea

\beq\ts
S_{5b}=\frac{1}{144}\,,
\eeq
\beq\ts
G_{5b}(g_1,g_2)=\frac{1}{9}N(N+2)^2g_1^4+\frac{4}{3}N(N+2)g_1^3g_2
+\frac{2}{3}N(N+8)g_1^2g_2^2+4Ng_1g_2^3+Ng_2^4\,.
\eeq

\bea
\bar{I}_{5c}
&\equiv&\ts
{\cal K}\bar{R}
\rule[-18pt]{0pt}{42pt}
\bpi(58,12)
\put(13,3){\circle{16}}
\put(45,3){\circle{16}}
\put(5,-5){\line(0,1){16}}
\put(25,-5){\oval(40,16)[b]}
\put(25,11){\oval(40,16)[t]}
\put(45,3){\oval(48,16)[l]}
\put(5,3){\circle*{4}}
\put(21,3){\circle*{4}}
\put(45,-5){\circle*{4}}
\put(45,11){\circle*{4}}
\epi
\nn\\
&=&\ts
\frac{1}{2}m^4{\cal K}\bar{R}\Bigg[
8\rule[-18pt]{0pt}{42pt}
\bpi(58,12)
\put(13,3){\circle{16}}
\put(45,3){\circle{16}}
\put(5,-5){\line(0,1){16}}
\put(25,-5){\oval(40,16)[b]}
\put(25,11){\oval(40,16)[t]}
\put(45,3){\oval(48,16)[l]}
\put(5,3){\circle*{4}}
\put(21,3){\circle*{4}}
\put(45,-5){\circle*{4}}
\put(45,11){\circle*{4}}
\put(51.9,-1){\circle*{4}}
\put(51.9,7){\circle*{4}}
\epi
+
4\rule[-18pt]{0pt}{42pt}
\bpi(58,12)
\put(13,3){\circle{16}}
\put(45,3){\circle{16}}
\put(5,-5){\line(0,1){16}}
\put(25,-5){\oval(40,16)[b]}
\put(25,11){\oval(40,16)[t]}
\put(45,3){\oval(48,16)[l]}
\put(5,3){\circle*{4}}
\put(21,3){\circle*{4}}
\put(45,-5){\circle*{4}}
\put(45,11){\circle*{4}}
\put(37,3){\circle*{4}}
\put(53,3){\circle*{4}}
\epi
+
32\rule[-18pt]{0pt}{42pt}
\bpi(58,12)
\put(13,3){\circle{16}}
\put(45,3){\circle{16}}
\put(5,-5){\line(0,1){16}}
\put(25,-5){\oval(40,16)[b]}
\put(25,11){\oval(40,16)[t]}
\put(45,3){\oval(48,16)[l]}
\put(5,3){\circle*{4}}
\put(21,3){\circle*{4}}
\put(45,-5){\circle*{4}}
\put(45,11){\circle*{4}}
\put(53,3){\circle*{4}}
\put(25,19){\circle*{4}}
\epi
+
8\rule[-18pt]{0pt}{42pt}
\bpi(58,12)
\put(13,3){\circle{16}}
\put(45,3){\circle{16}}
\put(5,-5){\line(0,1){16}}
\put(25,-5){\oval(40,16)[b]}
\put(25,11){\oval(40,16)[t]}
\put(45,3){\oval(48,16)[l]}
\put(5,3){\circle*{4}}
\put(21,3){\circle*{4}}
\put(45,-5){\circle*{4}}
\put(45,11){\circle*{4}}
\put(13,11){\circle*{4}}
\put(53,3){\circle*{4}}
\epi
\nn\\
&&\ts\hspace{45pt}
+
8\rule[-18pt]{0pt}{42pt}
\bpi(58,12)
\put(13,3){\circle{16}}
\put(45,3){\circle{16}}
\put(5,-5){\line(0,1){16}}
\put(25,-5){\oval(40,16)[b]}
\put(25,11){\oval(40,16)[t]}
\put(45,3){\oval(48,16)[l]}
\put(5,3){\circle*{4}}
\put(21,3){\circle*{4}}
\put(45,-5){\circle*{4}}
\put(45,11){\circle*{4}}
\put(33,11){\circle*{4}}
\put(25,19){\circle*{4}}
\epi
+
4\rule[-18pt]{0pt}{42pt}
\bpi(58,12)
\put(13,3){\circle{16}}
\put(45,3){\circle{16}}
\put(5,-5){\line(0,1){16}}
\put(25,-5){\oval(40,16)[b]}
\put(25,11){\oval(40,16)[t]}
\put(45,3){\oval(48,16)[l]}
\put(5,3){\circle*{4}}
\put(21,3){\circle*{4}}
\put(45,-5){\circle*{4}}
\put(45,11){\circle*{4}}
\put(33,-5){\circle*{4}}
\put(25,19){\circle*{4}}
\epi
+
8\rule[-18pt]{0pt}{42pt}
\bpi(58,12)
\put(13,3){\circle{16}}
\put(45,3){\circle{16}}
\put(5,-5){\line(0,1){16}}
\put(25,-5){\oval(40,16)[b]}
\put(25,11){\oval(40,16)[t]}
\put(45,3){\oval(48,16)[l]}
\put(5,3){\circle*{4}}
\put(21,3){\circle*{4}}
\put(45,-5){\circle*{4}}
\put(45,11){\circle*{4}}
\put(18.3,19){\circle*{4}}
\put(31.7,19){\circle*{4}}
\epi
\Bigg]
\nn\\
&=&\ts
\frac{m^4}{(4\pi)^{10}}
\bigg[\frac{96}{5\ep^5}-\frac{304}{5\ep^4}
+\frac{1}{\ep^3}\left(\frac{1592}{15}+\frac{192}{5}\ze(3)\right)
+\frac{1}{\ep^2}\left(-\frac{340}{3}+\frac{24}{5}\ze(3)
-\frac{96}{5}\ze(4)\right)
\nn\\
&&\ts\hspace{30pt}
+\frac{1}{\ep}\left(\frac{329}{8}-16\ze(3)+\frac{18}{5}\ze(4)
-\frac{64}{5}\ze(5)\right)\bigg]\,,
\eea

\beq\ts
S_{5c}=\frac{1}{32}\,,
\eeq
\beq\ts
G_{5c}(g_1,g_2)=\frac{1}{81}N(N+2)(5N+22)g_1^4+\frac{4}{27}N(5N+22)g_1^3g_2
+\frac{2}{9}N(N+26)g_1^2g_2^2+4Ng_1g_2^3+Ng_2^4\,.
\eeq

\bea
\bar{I}_{5d}
&\equiv&\ts
{\cal K}\bar{R}
\rule[-14pt]{0pt}{50pt}
\bpi(34,12)
\put(17,3){\circle{24}}
\put(17,23){\circle{16}}
\put(6.6,9){\line(1,0){20.8}}
\put(6.6,9){\line(3,-5){10.4}}
\put(27.4,9){\line(-3,-5){10.4}}
\put(6.6,9){\circle*{4}}
\put(27.4,9){\circle*{4}}
\put(17,-9){\circle*{4}}
\put(17,15){\circle*{4}}
\epi
=
\frac{1}{2}m^4{\cal K}\bar{R}\left[
4\rule[-14pt]{0pt}{48.8pt}
\bpi(36.85,12)
\put(19.85,3){\circle{24}}
\put(9.45,9){\line(1,0){20.8}}
\put(9.45,9){\line(3,-5){10.4}}
\put(30.25,9){\line(-3,-5){10.4}}
\put(9.45,9){\circle*{4}}
\put(30.25,9){\circle*{4}}
\put(19.85,-9){\circle*{4}}
\put(15.75,14.3){\circle*{4}}
\put(24.15,14.3){\circle*{4}}
\put(13,21.8){\circle{16}}
\put(10.25,29.3){\circle*{4}}
\epi
+
2\rule[-14pt]{0pt}{50pt}
\bpi(34,12)
\put(17,3){\circle{24}}
\put(17,23){\circle{16}}
\put(6.6,9){\line(1,0){20.8}}
\put(6.6,9){\line(3,-5){10.4}}
\put(27.4,9){\line(-3,-5){10.4}}
\put(6.6,9){\circle*{4}}
\put(27.4,9){\circle*{4}}
\put(17,-9){\circle*{4}}
\put(17,15){\circle*{4}}
\put(17,31){\circle*{4}}
\put(17,9){\circle*{4}}
\epi
+
8\rule[-14pt]{0pt}{50pt}
\bpi(34,12)
\put(17,3){\circle{24}}
\put(17,23){\circle{16}}
\put(6.6,9){\line(1,0){20.8}}
\put(6.6,9){\line(3,-5){10.4}}
\put(27.4,9){\line(-3,-5){10.4}}
\put(6.6,9){\circle*{4}}
\put(27.4,9){\circle*{4}}
\put(17,-9){\circle*{4}}
\put(17,15){\circle*{4}}
\put(17,31){\circle*{4}}
\put(6.6,-3){\circle*{4}}
\epi
\right]
\nn\\
&=&\ts
\frac{m^4}{(4\pi)^{10}}
\left[\frac{16}{\ep^5}-\frac{88}{3\ep^4}+\frac{28}{\ep^3}
+\frac{1}{\ep^2}\left(-\frac{25}{3}+4\ze(3)\right)\right]\,,
\eea

\beq\ts
S_{5d}=\frac{1}{16}\,,
\eeq
\beq\ts
G_{5d}(g,0)=\frac{1}{81}N(N+2)^2(N+8)g^4\,.
\eeq

\beq\ts
\bar{I}_{5e}
\equiv
{\cal K}\bar{R}
\rule[-14pt]{0pt}{47.3pt}
\bpi(46,12)
\put(23,3){\circle{24}}
\put(13,20.3){\circle{16}}
\put(33,20.3){\circle{16}}
\put(23,3){\oval(24,8)}
\put(11,3){\circle*{4}}
\put(35,3){\circle*{4}}
\put(17,13.4){\circle*{4}}
\put(29,13.4){\circle*{4}}
\epi
=
m^4{\cal K}\bar{R}
\rule[-14pt]{0pt}{47.3pt}
\bpi(46,12)
\put(23,3){\circle{24}}
\put(13,20.3){\circle{16}}
\put(33,20.3){\circle{16}}
\put(23,3){\oval(24,8)}
\put(11,3){\circle*{4}}
\put(35,3){\circle*{4}}
\put(17,13.4){\circle*{4}}
\put(29,13.4){\circle*{4}}
\put(9,27.25){\circle*{4}}
\put(37,27.25){\circle*{4}}
\epi
=
\frac{m^4}{(4\pi)^{10}}
\left(\frac{8}{3\ep^4}-\frac{3}{\ep^3}\right)\,,
\eeq

\beq\ts
S_{5e}=\frac{1}{48}\,,
\eeq
\beq\ts
G_{5e}(g,0)=\frac{1}{27}N(N+2)^3g^4\,.
\eeq

\beq\ts
\bar{I}_{5f}
\equiv
{\cal K}\bar{R}
\rule[-30pt]{0pt}{66pt}
\bpi(34,12)
\put(17,3){\circle{24}}
\put(17,-17){\circle{16}}
\put(17,23){\circle{16}}
\put(17,3){\oval(24,8)}
\put(5,3){\circle*{4}}
\put(17,-9){\circle*{4}}
\put(17,15){\circle*{4}}
\put(29,3){\circle*{4}}
\epi
=
m^4{\cal K}\bar{R}
\rule[-30pt]{0pt}{66pt}
\bpi(34,12)
\put(17,3){\circle{24}}
\put(17,-17){\circle{16}}
\put(17,23){\circle{16}}
\put(17,3){\oval(24,8)}
\put(5,3){\circle*{4}}
\put(17,-9){\circle*{4}}
\put(17,15){\circle*{4}}
\put(29,3){\circle*{4}}
\put(17,-25){\circle*{4}}
\put(17,31){\circle*{4}}
\epi
=
\frac{m^4}{(4\pi)^{10}}
\left(\frac{32}{3\ep^5}-\frac{32}{3\ep^4}+\frac{8}{3\ep^3}\right)\,,
\eeq

\beq\ts
S_{5f}=\frac{1}{32}\,,
\eeq
\beq\ts
G_{5f}(g,0)=\frac{1}{27}N(N+2)^3g^4\,.
\eeq

\beq\ts
\bar{I}_{5g}
\equiv
{\cal K}\bar{R}
\rule[-14pt]{0pt}{66pt}
\bpi(34,12)
\put(17,3){\circle{24}}
\put(17,23){\circle{16}}
\put(17,39){\circle{16}}
\put(17,3){\oval(24,8)}
\put(5,3){\circle*{4}}
\put(17,15){\circle*{4}}
\put(17,31){\circle*{4}}
\put(29,3){\circle*{4}}
\epi
=
\frac{1}{2}m^4{\cal K}\bar{R}\left[
4\rule[-14pt]{0pt}{61.15pt}
\bpi(54,12)
\put(31,3){\circle{24}}
\put(21,20.3){\circle{16}}
\put(13,34.15){\circle{16}}
\put(31,3){\oval(24,8)}
\put(19,3){\circle*{4}}
\put(43,3){\circle*{4}}
\put(25,13.4){\circle*{4}}
\put(37,13.4){\circle*{4}}
\put(17,27.25){\circle*{4}}
\put(9,41.1){\circle*{4}}
\epi
+
6\rule[-14pt]{0pt}{66pt}
\bpi(34,12)
\put(17,3){\circle{24}}
\put(17,23){\circle{16}}
\put(17,39){\circle{16}}
\put(17,3){\oval(24,8)}
\put(5,3){\circle*{4}}
\put(17,15){\circle*{4}}
\put(17,31){\circle*{4}}
\put(29,3){\circle*{4}}
\put(17,47){\circle*{4}}
\put(17,-9){\circle*{4}}
\epi
\right]
=
\frac{m^4}{(4\pi)^{10}}
\left(\frac{32}{\ep^5}-\frac{80}{3\ep^4}+\frac{2}{\ep^3}\right)\,,
\eeq

\beq\ts
S_{5g}=\frac{1}{48}\,,
\eeq
\beq\ts
G_{5g}(g,0)=\frac{1}{27}N(N+2)^3g^4\,.
\eeq

\beq\ts
\bar{I}_{5h}
\equiv
{\cal K}\bar{R}
\rule[-10pt]{0pt}{26pt}
\bpi(90,12)
\put(13,3){\circle{16}}
\put(29,3){\circle{16}}
\put(45,3){\circle{16}}
\put(61,3){\circle{16}}
\put(77,3){\circle{16}}
\put(21,3){\circle*{4}}
\put(37,3){\circle*{4}}
\put(53,3){\circle*{4}}
\put(69,3){\circle*{4}}
\epi
=
m^4{\cal K}\bar{R}
\rule[-10pt]{0pt}{26pt}
\bpi(90,12)
\put(13,3){\circle{16}}
\put(29,3){\circle{16}}
\put(45,3){\circle{16}}
\put(61,3){\circle{16}}
\put(77,3){\circle{16}}
\put(5,3){\circle*{4}}
\put(21,3){\circle*{4}}
\put(37,3){\circle*{4}}
\put(53,3){\circle*{4}}
\put(69,3){\circle*{4}}
\put(85,3){\circle*{4}}
\epi
=
\frac{32m^4}{(4\pi)^{10}\ep^5}\,,
\eeq

\beq\ts
S_{5h}=\frac{1}{64}\,,
\eeq
\beq\ts
G_{5h}(g,0)=\frac{1}{81}N(N+2)^4g^4\,.
\eeq

\beq\ts
\bar{I}_{5i}
\equiv
{\cal K}\bar{R}
\rule[-23.85pt]{0pt}{53.7pt}
\bpi(66,12)
\put(13,3){\circle{16}}
\put(29,3){\circle{16}}
\put(45,3){\circle{16}}
\put(53,-10.85){\circle{16}}
\put(53,16.85){\circle{16}}
\put(21,3){\circle*{4}}
\put(37,3){\circle*{4}}
\put(49,-3.9){\circle*{4}}
\put(49,9.9){\circle*{4}}
\epi
=
0\,,
\eeq

\beq\ts
S_{5i}=\frac{1}{32}\,,
\eeq
\beq\ts
G_{5i}(g,0)=\frac{1}{81}N(N+2)^4g^4\,.
\eeq

\beq\ts
\bar{I}_{5j}
\equiv
{\cal K}\bar{R}
\rule[-26pt]{0pt}{58pt}
\bpi(58,12)
\put(13,3){\circle{16}}
\put(29,-13){\circle{16}}
\put(29,3){\circle{16}}
\put(29,19){\circle{16}}
\put(45,3){\circle{16}}
\put(21,3){\circle*{4}}
\put(29,-5){\circle*{4}}
\put(29,11){\circle*{4}}
\put(37,3){\circle*{4}}
\epi
=
0\,,
\eeq

\beq\ts
S_{5j}=\frac{1}{128}\,,
\eeq
\beq\ts
G_{5j}(g,0)=\frac{1}{81}N(N+2)^4g^4\,.
\eeq

\setlength{\baselineskip}{14pt}

\end{document}